\documentclass[lettersize,journal]{IEEEtran}
\usepackage{amsmath,amsfonts}
\usepackage{amssymb}
\usepackage{algorithm}
\usepackage{algorithmicx}
\usepackage{array}
\usepackage[caption=false,font=footnotesize,labelfont=rm,textfont=rm]{subfig}
\usepackage{textcomp}
\usepackage{stfloats}
\usepackage{url}
\usepackage{verbatim}
\usepackage{graphicx}
\usepackage{cite}
\usepackage{ulem}
\usepackage{threeparttable}
\usepackage{newtxmath} 
\usepackage{enumitem}  

\usepackage[
pdfauthor={derajan},
pdftitle={How to do this},
pdfstartview=XYZ,
bookmarks=true,
colorlinks=true,
linkcolor=blue,
urlcolor=blue,
citecolor=blue,
pdftex,
bookmarks=true,
linktocpage=true,   
hyperindex=true
]{hyperref}

\begin{document}

\title{Dim and Small Target Detection for Drone Broadcast Frames Based on Time-Frequency Analysis}

\author{Jie Li,
	Jing Li,~\IEEEmembership{Member,~IEEE,}
	Lu Lv,~\IEEEmembership{Member,~IEEE,}
	Zhanyu Ju,
	Fengkui Gong,~\IEEEmembership{Member,~IEEE,}
		\thanks{ \textit{(Corresponding author: Jing Li.)}}
		\thanks{Jie Li, Jing Li, Zhanyu Ju, and Fengkui Gong are with the State Key Laboratory of Integrated Services Network, Xidian University, Xi'an, Shaanxi 710071, China (e-mail: lijie\_372@stu.xidian.edu.cn; jli@xidian.edu.cn; juzhanyu@stu.xidian.edu.cn; fkgong@xidian.edu.cn).}
		\thanks{Lu Lv is with the School of Telecommunications Engineering, Xidian University, Xi'an 710071, China, and also with the Shaanxi Key Laboratory of Information Communication Network and Security, Xi'an University of Posts \& Telecommunications, Xi'an 710121, China (e-mail: lulv@xidian.edu.cn).}
}

\maketitle

\begin{abstract}
We propose a dim and small target detection algorithm for drone broadcast frames based on the time-frequency analysis of communication protocol. Specifically, by analyzing modulation parameters and frame structures, the prior knowledge of transmission frequency, signal bandwidth, Zadoff-Chu (ZC) sequences, and frame length of drone broadcast frames is established. The RF signals are processed through the designed filter banks, and the frequency domain parameters of bounding boxes generated by the detector are corrected with transmission frequency and signal bandwidth. Given the remarkable correlation characteristics of ZC sequences, the frequency domain parameters of bounding boxes with low confidence scores are corrected based on ZC sequences and frame length, which improves the detection accuracy of dim targets under low signal-to noise ratio situations. Besides, a segmented energy refinement method is applied to mitigate the deviation caused by interference signals with high energy strength, which ulteriorly corrects the time domain detection parameters for dim targets. As the sampling duration increases, the detection speed improves while the detection accuracy of broadcast frames termed as small targets decreases. The trade-off between detection accuracy and speed versus sampling duration is established, which helps to meet  different drone regulation requirements. Simulation results demonstrate that the proposed algorithm improves the evaluation metrics by 2.27\% compared to existing algorithms. The proposed algorithm also performs strong robustness under varying flight distances, diverse types of environment noise, and different flight visual environment. Besides, the broadcast frame decoding results indicate that 97.30\% accuracy of RID has been achieved.
\end{abstract}


\section{Introduction}
\label{sec1}
\IEEEPARstart{T}{he} low-altitude economy, which is rapidly developing, refers to an economic model focused on the flight activities of both manned and unmanned aerial vehicles within airspace ranging from the ground up to 1000 m\textcolor{blue}{\cite{ref1}}. The low-altitude system composed of low-cost and highly flexible drones, serves as the foundational infrastructure for the low-altitude industrialization and a key component of the low-altitude economy\textcolor{blue}{\cite{ref2}}. While the widely deployed drones have enhanced economic growth across various areas including logistics, agriculture, and tourism, they have also introduced new security threats, such as unauthorized malicious flights and eavedropping attacks\textcolor{blue}{\cite{ref3}},\textcolor{blue}{\cite{ref4}},\textcolor{blue}{\cite{ref5}},\textcolor{blue}{\cite{ref6}}. Although many countries and organizations have issued policies and rules for drone registration, operator training, and flight authorization, there remains a lack of effective methods for drones remote identification (RID)\textcolor{blue}{\cite{ref7}}. Existing works mainly focus on the detection and classification of drones, which have neglected the object tracking and trajectory prediction. Specifically, the radar echo signals\textcolor{blue}{\cite{ref8}},\textcolor{blue}{\cite{ref9}}, visual images\textcolor{blue}{\cite{ref10}}, acoustic signals\textcolor{blue}{\cite{ref11}}, and radio-frequency (RF) signals\textcolor{blue}{\cite{ref12}},\textcolor{blue}{\cite{ref13}} are commonly utilized to detect and classify the types and operation modes of drones. However, effective drone regulation requires accurate tracking and prediction of drone positions and trajectories, as mere classification are insufficient for further analyzing drone intents and implementing countermeasures.

Due to the RID regulations, drones are required to regularly broadcast their identity, current location, and details of the related ground control station, with no mandatory encryption nor authentication\textcolor{blue}{\cite{ref7}}. Thus, once the broadcast frames are detected with time-frequency position, the plain text can be obtained to track and predict drones. The transmission frequency of broadcast frames spans a 100 MHz bandwidth across the 2.4 GHz and 5.8 GHz bands, occurring at an average interval of 0.6\(\sim\)3 s, with a duration of 572 or 643 \(\upmu\)s\textcolor{blue}{\cite{ref14}},\textcolor{blue}{\cite{ref15}}, which make detection termed as small targets particularly challenging. Furthermore, the presence of uplink control signals, downlink video transmission signals, unknown interference, and background noise\textcolor{blue}{\cite{ref16}} makes improving the detection accuracy of broadcast frames termed as dim targets critically important.

\subsection{Related Works}

The radar-based RID algorithms utilized the characteristics of radar echo signals to detect drones, but the micro drones weighting less than 250 grams may evade detection due to their extremely small radar cross-section\textcolor{blue}{\cite{ref8}},\textcolor{blue}{\cite{ref9}}. The visual-based RID algorithms adopted deep learning such as You Only Look Once (YOLO)\textcolor{blue}{\cite{ref17}} and faster region-based convolutional neural networks\textcolor{blue}{\cite{ref18}} to analyze images and videos for identifying drones. However, its robustness was limited under various weather conditions\textcolor{blue}{\cite{ref19}}, diverse resolutions\textcolor{blue}{\cite{ref20}}, different flight distances\textcolor{blue}{\cite{ref21}}, and occlusion scenarios\textcolor{blue}{\cite{ref10}}. The acoustic-based RID algorithms captured the sound of propellers based on microphone arrays to identify drones, which were susceptible to environment noise, flight distance, and the presence of noise reduction technology\textcolor{blue}{\cite{ref22}},\textcolor{blue}{\cite{ref23}}. The RF-based RID algorithms analyzed the RF signals from drones and remote controls, which can detect the micro drones without the constraints of weather and flight distance.

Since the drone signals will change the electromagnetic environments, the variance of higher-order cumulants can be computed for the received signals to detect the existence of drones\textcolor{blue}{\cite{ref24}}. But this algorithm exhibited poor robustness in various flight distance and signal-to noise ratio (SNR) scenarios, which achieved only 10\% and 40\% detection accuracy at a flight distance of 35 m and an SNR of 0 dB, respectively. Under a given drone coordinates and channel conditions, the received signal strength can be utilized to detect drones, which achieved 50\% detection accuracy at a flight distance of 200 meters and an SNR of 6 dB without any interference signals\textcolor{blue}{\cite{ref25}}. The uplink frequency hopping spread spectrum (FHSS) control signals can induce regular and detectable electromagnetic radiation, which can be extracted by the energy fusion from a series of harmonic components\textcolor{blue}{\cite{ref26}}. In the process of achieving 95\% detection accuracy at a flight distance of 60 m, 4.5 s\textcolor{blue}{\cite{ref26}}, 6.8 s\textcolor{blue}{\cite{ref11}}, and 8.5 s\textcolor{blue}{\cite{ref12}} were spent, respectively. Besides, the emerging integrated sensing and communication technology was expected to provide promising schemes for detecting drones' number, position, and velocity\textcolor{blue}{\cite{ref27}},\textcolor{blue}{\cite{ref28}}, which can be achieved by the individual base station, connected drones, and multiple base stations \textcolor{blue}{\cite{ref29}},\textcolor{blue}{\cite{ref30}}.

Due to the fact that the uplink FHSS control signals had higher amplitude than the downlink orthogonal frequency division multiplexing (OFDM) video transmission signals, the in-phase/quadrature (I/Q) sequences sampled in muffler chamber were converted into time-frequency images (TFI) with modified FHSS features, which were utilized to achieve drones RID\textcolor{blue}{\cite{ref31}}. Since malicious drones can turn off the downlink OFDM video transmission signals, the uplink FHSS control signals were more indispensable and reliable to be utilized for identifying drones\textcolor{blue}{\cite{ref32}}. However, the bandwidth and duration of FHSS signals is often 0.4\(\sim\)5 MHz and 0.5\(\sim\)5 ms, respectively, which can be seen as the small targets in TFIs with a bandwidth of 100 MHz and a duration typically exceeding 20 ms. Besides, the multiple-signal coexisting and low SNR scenarios make the FHSS signals difficult to detect termed as dim targets\textcolor{blue}{\cite{ref33}}.

The edge and corner gradient features were enhanced to improve the robustness  against 6 typical noises, then the positional and textural features were integrated by the feature pyramid network\textcolor{blue}{\cite{ref34}} to improve the detection and classification accuracy on small targets\textcolor{blue}{\cite{ref32}}. Since the small targets with approximate isotropy can be  modeled by the 2-dimensional Gaussian model, the second-order direction derivative feature with convolution operation was utilized to represent the targets' gradient strength distribution, which can suppress the background pixels and enhance the small targets\textcolor{blue}{\cite{ref35}}. In order to pay more attention on signals with lower SNR, a nonlinear transformation function was adopted to expand the dynamic range of low grayscale values, while the range of high grayscale values remained relatively unchanged\textcolor{blue}{\cite{ref36}}. This algorithm increased the contrast between signals and background noise, and improved the signal identification accuracy for SNR above 0 dB. Besides, exploring the loss function with higher sensitivity to scale and location\textcolor{blue}{\cite{ref37}}, the anchor-free signal detector based on multi-grained time-frequency localization strategy\textcolor{blue}{\cite{ref38}}, the enhancement of feature representation\textcolor{blue}{\cite{ref39}}, and the corresponding attention mechanisms associated with different levels of features\textcolor{blue}{\cite{ref40}} can further improve the detection accuracy for dim and small targets.

Motivated by the advances in computer vision, several studies had explored reconstructing cleaner TFIs from TFI consisted of overlapping signals under low SNR situation. The decoder will extract features from the masked TFIs, then the pre-trained decoder reconstructed TFIs to eliminate the phenomenon of signal overlapping, which also improved the detection accuracy of the dim target signals\textcolor{blue}{\cite{ref41}}. The unsupervised denoising and decomposited enhanced network was proposed to achieve the removal of Gaussian noise, wireless fidelity (Wi-Fi), and Bluetooth signals, which
significantly improved the quality of downlink OFDM video transmission signals\textcolor{blue}{\cite{ref42}}.
TFIs with noise can be regarded as the hazy or low resolution version of clean and high quality TFIs, which can be reconstructed to obtain the high resolution TFIs by diffusion model\textcolor{blue}{\cite{ref43}} or teacher-student model\textcolor{blue}{\cite{ref44}}.

\subsection{Motivations and Contributions}
Despite previous works have explored various drone detection and classification algorithms, several challenges remain in achieving effective drone regulation, as follows:

\begin{itemize}
	\item{Insufficient research on leveraging drone broadcast frames for RID. Unlike non-protocol-based algorithms, broadcast frames convey plain text information as required by communication protocols, providing a promising choice for drone tracking and trajectory prediction. Thus, the tracking and prediction task can be transformed into the detection and decoding of broadcast frames.
	}
	\item{Superiority of data-level over model-level decoupling for broadcast frame detection termed as dim and small targets. Traditional image enhancement methods, such as dehazing or denoising, used in dim and small target detection are ineffective for broadcast frames. This is due to the similar characteristics between broadcast frames and interference signals in TFIs. Pixel-level image enhancements may not consistently and reliably improve semantic-level detection, especially under low-SNR and signal overlapping conditions. In contrast, leveraging prior knowledge of communication protocol, such as modulation parameters and frame structures, enables data-level decoupling and detection performance enhancement.
	}
	\item{Bounding box deviations in TFIs under practical environments. Under low SNR situations, broadcast frames may be obscured by noise, causing bounding boxes to deviate significantly from the ground truth. In the presence of high-energy interference or overlapping signals, bounding boxes may shift to the time-frequency regions of the interference, often with low confidence scores. Since bounding boxes with high confidences are typically more accurate, it becomes necessary to correct the time-frequency parameters of those with low confidence to improve overall detection precision.
	}
	\item{Trade-off between detection accuracy and speed in broadcast frame detection. Unlike natural image-based object detection tasks, TFIs used for drone RID must retain a bandwidth of approximately 100 MHz to conform to protocol constraints. While different sampling durations influence whether broadcast frames appear as dim and small targets in TFIs, the inference speed (e.g., frame per second (FPS)\textcolor{blue}{\cite{ref45}} or latency\textcolor{blue}{\cite{ref11}},\textcolor{blue}{\cite{ref12}},\textcolor{blue}{\cite{ref16}},\textcolor{blue}{\cite{ref26}}) of the pre-trained detector remains fixed. Thus, optimizing the trade-off between sampling duration and detection accuracy is crucial for meeting real-time RID requirements.
	}
\end{itemize}

To address the above challenges, we propose a time-frequency analysis-based dim and small target detection algorithm for drone broadcast frames, utilizing outdoor practical sampling datasets named DroneRFa\textcolor{blue}{\cite{ref46}} and DroneRFb-DIR\textcolor{blue}{\cite{ref47}}. Specifically, the YOLOv7 network\textcolor{blue}{\cite{ref48}} is leveraged to identify and locate broadcast frames within TFIs, while the prior knowledge of modulation parameters and frame structures is derived from communication protocols. The frequency domain parameters of bounding boxes are corrected by the transmission frequency and signal bandwidth, while the time-domain parameters are modified based on the cross-correlation results of Zadoff-Chu (ZC) sequences and frame length. Additionally, the impact of interference signals with high energy strength on cross-correlation results is mitigated by adopting a segmented energy refinement method. The contributions of this paper are summarized as follows.

\begin{itemize}
	\item{Prior knowledge of the modulation parameters and frame structures of drone broadcast frames is systematically established. The potential transmission frequencies of broadcast frames, spanning the 100 MHz sampling bandwidth at 2.4 GHz and 5.8 GHz as regulated by communication protocols, are statistically summarized. Additionally, the ZC sequences adopted in broadcast frames are identified based on the estimation of signal bandwidth and the number of subcarriers.
	}
	\item{A data-level decoupling algorithm is proposed for bounding box parameter correction based on protocol prior knowledge. Three sets of filter banks, specifically designed for 2.4 GHz and 5.8 GHz based on the transmission frequency set and signal bandwidth, are applied before detection. For bounding boxes in the filtered TFIs, frequency domain parameters are corrected according to the fixed signal bandwidth. The peak's index of the ZC sequence cross-correlation results is utilized to determine the beginning location of broadcast frames, while the time domain parameters of low-confidence bounding boxes are corrected based on the prior knowledge of frame length. Furthermore, to mitigate the impact of interference signals with high energy strength on cross-correlation results, a segmented energy refinement module is deployed.
	}
	\item{The trade-off between detection accuracy and speed is quantitatively analyzed. Although the YOLOv7 detector's FPS remains fixed, detection latency varies depending on the sampling duration due to time domain parameters correction overhead. Detection accuracy fluctuates since sampling duration affects whether broadcast frames appear as dim and small targets in TFIs. By evaluating broadcast frame detection performance across different sampling durations, the proposed algorithm provides flexibility for adapting to various drone regulation scenarios.
	}
	\item{A decoding algorithm for broadcast frames has been proposed. Based on the time-frequency parameters of the bounding boxes, cyclic prefix (CP) and ZC sequences are utilized for time and frequency synchronization, and bitstream is obtained. Based on the analysis of communication protocols, the descrambling and Turbo decoding are performed, and the payload information is recovered according to the regulated message format.
	}
	\item{Numerical results demonstrate that the proposed algorithms effectively achieve drone RID across different channel conditions and drone types. Specifically: 1) The proposed algorithm achieves at least 3\%, 1.6\%, and 2.4\% improvement in intersection over union (IoU), precision, and recall over existing baselines, confirming the efficacy of time-frequency analysis of communication protocols. 2) The algorithms exhibit strong robustness across variable flight distances, visual environments, and noise types. 3) The proposed algorithms achieve the best trade-off between detection accuracy and latency, ensuring feasibility for practical deployment. 4) The decoding results for different drone broadcast frames indicate that the proposed algorithm achieve effective RID.
	}
\end{itemize}
\subsection{Outline and Notations}
The signal analysis is presented in Section \ref{sec2}. The details of drone broadcast frames detection and decoding algorithm are discussed in Section \ref{sec3} and Section \ref{sec4}, respectively. Simulation evaluation is presented in Section \ref{sec5}. And this paper is summarized in Section \ref{sec6}.

\section{Signal Analysis}
\label{sec2}
\subsection{Signal Dateset}

Both datasets are collected in an urban outdoor environment, in addition to drone downlink broadcast frames, uplink FHSS control signals, and downlink OFDM video transmission signals, the RF signals also contain Bluetooth and Wi-Fi signals as interference. Signals in DroneRFa are sampled in the 2.4 GHz and 5.8 GHz bands with sampling rate \(f_s=100\) MHz, while signals in DroneRFb-DIR are sampled in the 2.4 GHz band with \(f_s=80\) MHz.

In DroneRFa, drones transition from standby mode to flight mode, and RF signals are collected at flight distances ranging from 20 to 150 m, with one communication frequency switch occurring during the sampling process. The RF signals for each drone type are collected from a single drone, with a total of seven types of drones selected, all transmitting broadcast frames.

In DroneRFb-DIR, drones transition from standby mode to flight mode while maintaining a flight distance of 10 m and communication frequency of 2.4 GHz. Due to building obstructions, drone signals are received in both line-of-sight (LOS) and non-LOS flight environments, resulting in a SNR variation of up to 20 dB. Each drone type includes three drones, with a total of four types of drones selected, all transmitting broadcast frames.

The interpretation of the binary labels is provided in Table \ref{tab:table1}. Furthermore, to more realistically evaluate the robustness of the proposed algorithm in practical scenarios, additive white Gaussian noise (AWGN), Rayleigh noise, Gamma noise, and impulse noise are introduced into the RF signals, forming the dataset used in this paper.

\begin{table}[!t]   
	\caption{Explanation of the Dataset's Labels}  
	\label{tab:table1} 
	\centering
	\begin{threeparttable}
		\begin{tabular}{|c|c|c|c|c|}   
			\hline   \textbf{Label1} & \textbf{Labe2} & \textbf{Labe3} & \textbf{Labe4} & \textbf{Drone Types} \\   
			\hline   T0000 & D00 & S00 S01 & L0\(\sim\)L5 & DJI Air 2S \\ 
			\hline   T0001 & D00 & S00 & L0\(\sim\)L5 & DJI Mini 3 Pro \\  
			\hline   T0010 & D00 & S00 & L0\(\sim\)L5 & DJI Mavic Pro \\
			\hline   T0011 & D00 D01 D10 & S00 & L0\(\sim\)L5 & DJI Mini 2 \\ 
			\hline   T0100 & D00 & S00 & L0\(\sim\)L5 & DJI Mavic 3 \\ 
			\hline   T0101 & D00 & S00 & L0\(\sim\)L5 & DJI MATRICE 300 \\ 
			\hline   T0110 & D00 & S00 & L0\(\sim\)L5 & DJI MATRICE 30T \\
			\hline   T0111 & D00 & S00 S01 & L0\(\sim\)L5 & DJI Mavic 3 Pro \\
			\hline   T1000 & D00 & S00 S01 & L0\(\sim\)L5 & DJI Mini 2 SE \\
			\hline   T1001 & D00 & S00 S01 & L0\(\sim\)L5 & DJI Mini 3 \\
			\hline   
		\end{tabular}
		\begin{tablenotes}
			\footnotesize
			\item[1] D00, D01, and D10 denote the flight distance range of 20\(\sim\)40, 40\(\sim\)80, and 80\(\sim\)150 m, respectively.
			\item[2] L0, L1, L2, L3, L4, and L5 denote the sampling duration range of 1, 2, 5, 10, 20, and 50 ms, respectively.
			\item[3] S00 and S01 denote LOS and non-LOS flight environment, respectively.
		\end{tablenotes}            
	\end{threeparttable}  
\end{table}

\subsection{Frame Structure}
Let \(x(l)\) denote the received RF complex sequences with \(f_s=100\) MHz, where \(l = 0, 1, \ldots, L-1\) with \(L\) being the number of samples. The TFI matrix \(\mathbf{X}\) of \(x(l)\) can be computed by the short-time Fourier transform\textcolor{blue}{\cite{ref49}}, and the value of \(t\text{-th}\) time index and \(f\text{-th}\) frequency index is given by
\begin{equation}
	\label{Eq1}
	{{\mathbf{X}}_{t,f}}=\left| \sum\limits_{u =-\infty }^{+\infty }{x(u){h}(u-t){{\text{e}}^{\frac{-\text{j}2\pi\!fu}{U} }}} \right|,
\end{equation}
where \(h(u-t)\) denotes the window function located at \(t\text{-th}\) time index, the window length \(U\) is equal to the fast Fourier transform size. For signal \(x(l)\) of length \(L\), the computational complexity of generating one TFI is \(\mathcal{O}( L\log U )\).
\begin{figure}[!t]
	\centering
	\includegraphics[width=2.5in]{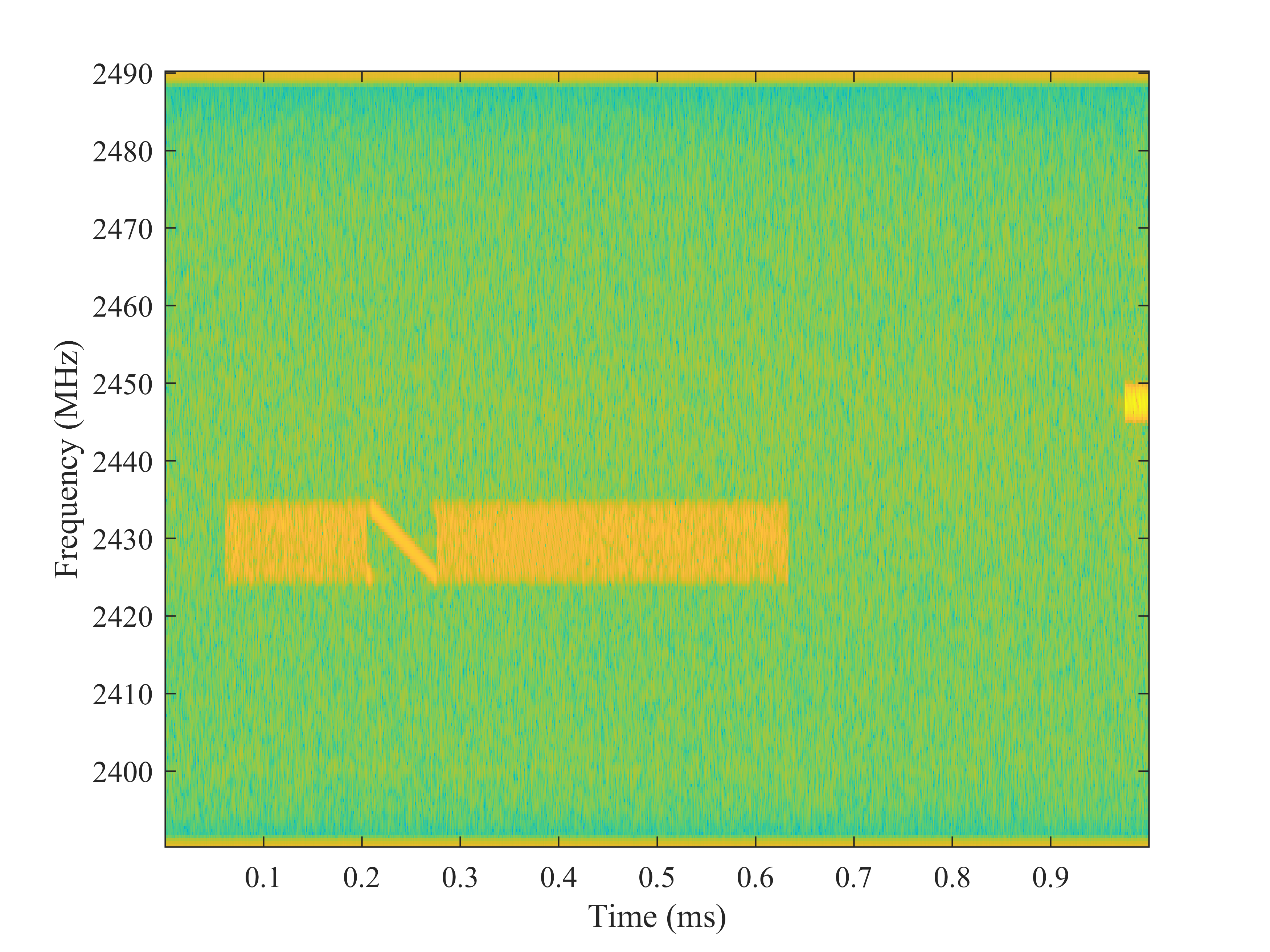}
	\caption{TFI of one broadcast frame of T0001D00S00L0.}
	\label{Fig_1}
\end{figure}

Fig. \ref{Fig_1} shows the TFI of one broadcast frame for \(L=10^5\) samples, it can be seen that the broadcast frame contains 8 OFDM symbols. It can also be observed that a broadcast frame consistently contains two distinct ZC sequences, while the number of the remaining data symbols is either 6 or 7, which varies depending on the type of drones. In order to obtain more detailed and useful prior knowledge, it is essential to estimate the signal bandwidth \(B\) and the number of subcarriers \(N\).

\subsection{Estimation of Modulation Parameters}
\begin{enumerate}[leftmargin=0pt, itemindent=2pc, listparindent=\parindent]
	\item{\textit{Signal Bandwidth}}:
	Let \(w(n)\) denote a window with length \(H\), the received signal \(x(l)\) can be divided into \(K\) overlapping sequences with offset \(D\), which can be expressed as
	\begin{equation}
		\label{Eq2}
		{{x}_{i}}(l)=x(iD+l)w(n), 0 \le i \le K-1,
	\end{equation}
	where \(D\) is set as \(\frac{H}{2}\) to reduce the estimation variance in Welch method\textcolor{blue}{\cite{ref50}}. The power spectral of \(x(l)\) can be obtained by
	\begin{equation}
		\label{Eq3}
		{{R}_{x}}({\text{e}^{\text{j}\omega }})=\frac{1}{KH}{{\left| \sum\limits_{l=0}^{L-1}{{{x}_{i}}(l){\text{e}^{-\text{j}\omega l}}} \right|}^{2}}.
	\end{equation}
		
	Let the maximum power of \({{R}_{x}}({\text{e}^{\text{j}\omega }})\) be \(P_{\max}\), the frequencies at which \({{R}_{x}}({\text{e}^{\text{j}\omega }}) = \frac{P_{\max}}{2} \) are denoted as \(F_{\text{lower}}\) and \(F_{\text{upper}}\), then the estimation of bandwidth \(\hat{B}_u\) can be expressed as
	\begin{equation}
		\label{Eq4}
		\hat{B}_u=F_{\text{upper}}-F_{\text{lower}}.
	\end{equation}
		
	However, as regulated in long term evolution (LTE)\textcolor{blue}{\cite{ref51}}, only \(N_u\) used subcarriers in an OFDM symbol are utilized for data transmission, there are \(N_v\) subcarriers are designed as virtual subcarriers to protect the signal frequency band, which are shown in Fig. \ref{Fig_2}. Thus, the estimated signal bandwidth \(\hat{B}_u\) in \textcolor{blue}{(\ref{Eq4})} corresponds only to \(N_u+1\) subcarriers, while the signal bandwidth \(\hat{B}\) for \(N\) subacrriers can be calculated by
	\begin{figure}[!t]
		\centering
		\includegraphics[width=2.5in]{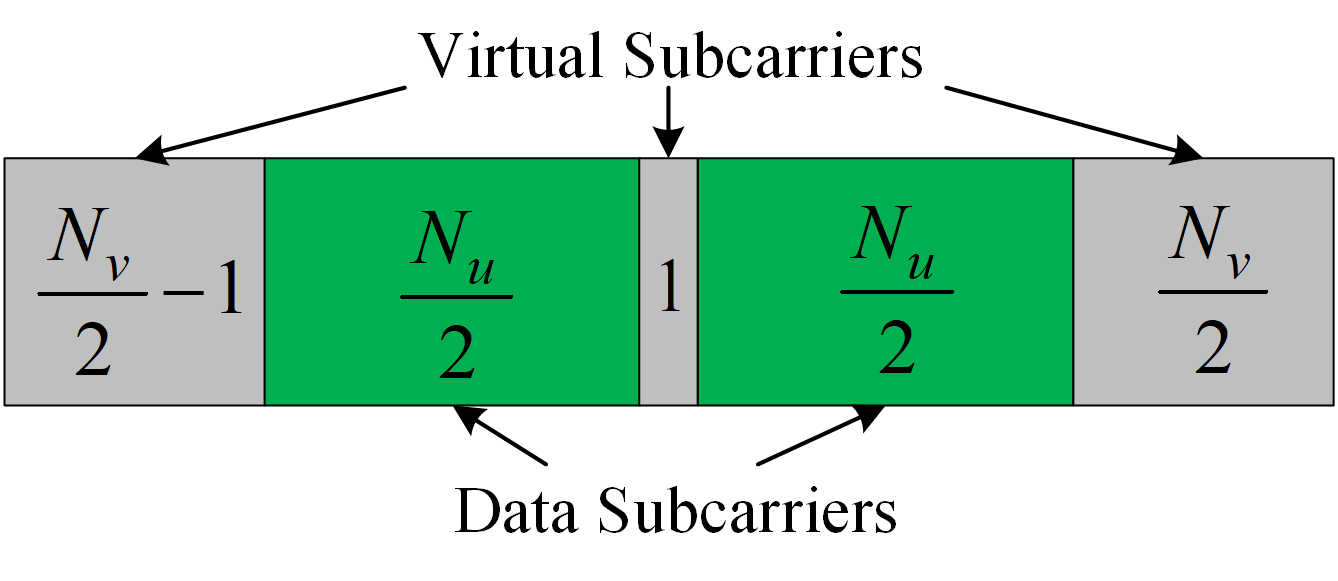}
		\caption{The distribution of \(\tiny{N}\) subcarriers for one OFDM symbol.}
		\label{Fig_2}
	\end{figure}
	\begin{equation}
		\label{Eq5}
		\hat{B}=\frac{\hat{B}_uN}{N_u+1}.
	\end{equation}
		
	Specifically, for the estimated signal bandwidth \(\hat{B}_u=9.015\) MHz, given the corresponding subcarrier interval of 15 KHz, it can be calculated that there exists \(N_u=600\) used subcarriers and 1 direct-current subcarrier. Then, it can also be inferred that there are \(N_v-1\) virtual subcarriers on both sides of the used subcarriers, with 212 on one side and 211 on the other side. For OFDM signals with different \(N\) and \(B\), the bandwidth estimation is performed after sampling with \(f_s\), and the simulation results are shown in Fig. \ref{Fig_3}. The parts per million (PPM) can be computed by \(\frac{\left|B-\hat{B}\right|}{10^6}\), and it can be observed that the estimation error of \(B\) does not exceed 120 KHz for commonly used \(N=1024\) and 2048. Moreover, since OFDM signals have a fixed set of \(B\) in MHz, such an error has no impact on acquiring prior knowledge of the signal bandwidth of broadcast frames.
	\begin{figure}[!t]
		\centering
		\includegraphics[width=2.5in]{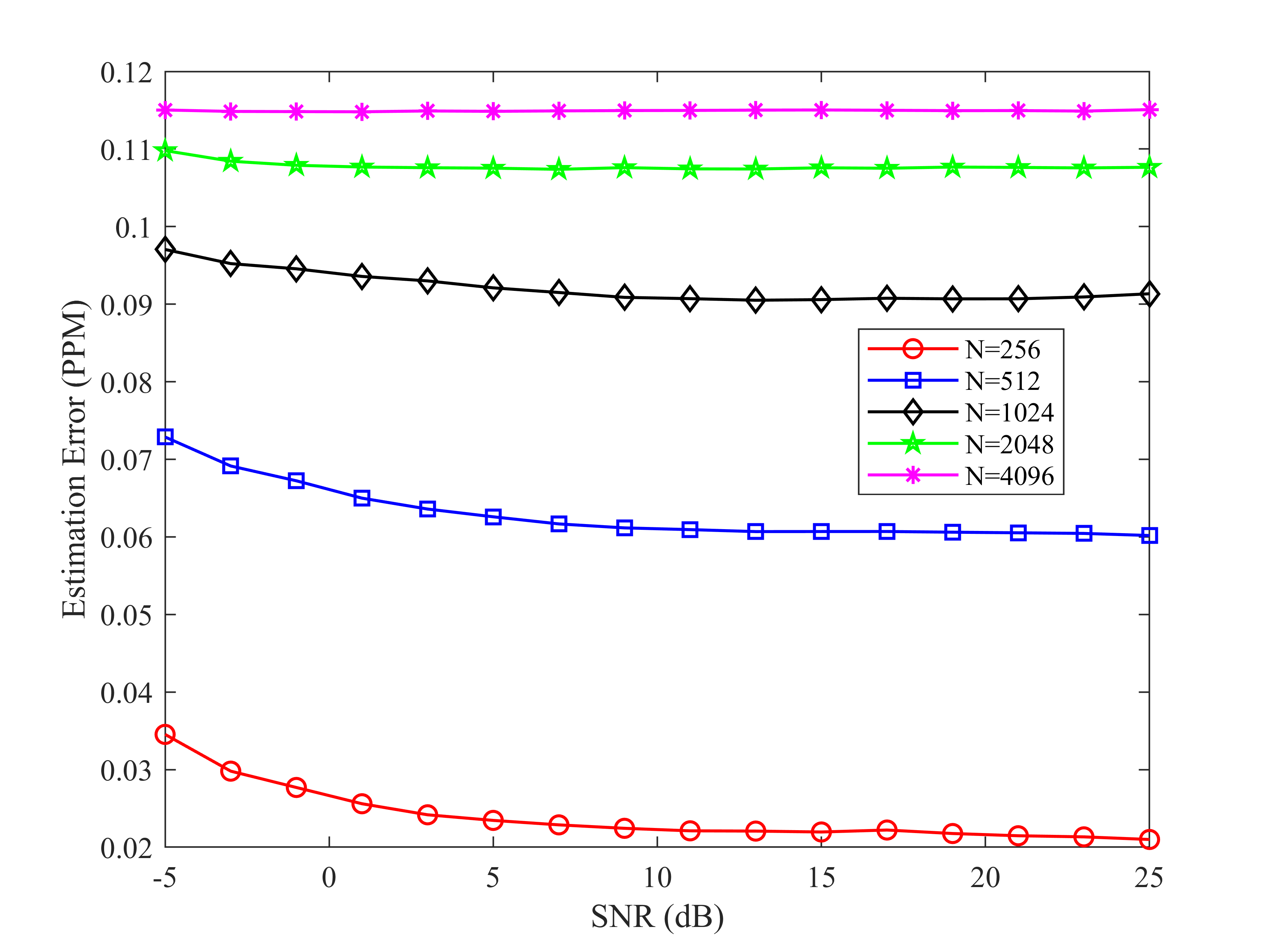}
		\caption{The estimation error of \(\tiny{\hat{B}}\) versus SNR for \(\tiny{f_s=100}\) MHz.}
		\label{Fig_3}
	\end{figure}
	
	\item{\textit{Number of Subcarriers}}:
	Since OFDM signals add CP before each symbol to eliminate inter-symbol interference and inter-carrier interference caused by multipath propagation, the auto-correlation characteristic of CP can be utilized to estimate the number of subcarriers \(N\). The auto-correlation results for \(x(l)\) with a window length \(L_a\) can be obtained by
	\begin{equation}
	\label{Eq6}
		\gamma(n)=\left| \sum\limits_{k=1}^{L_a}{x(k){{x}^{*}}(k+n)} \right|,n=0,1,\ldots ,L-L_a-1.
	\end{equation}
	
	Due to the fact that CP is copied from the tail segment of OFDM symbol with a length of \(N_{\text{cp}}\), \(\gamma(n)\) will exhibit a peak at index \(n^*=\frac{N{{f}_{s}}}{B}\) for \(L_a\geq\frac{N_{\text{cp}}{{f}_{s}}}{B}\) and \(L\geq\frac{N{{f}_{s}}}{B}\). Furthermore, the estimated signal bandwidth \(\hat{B}\) and the number of subcarriers \(\hat{N}\) can be mutually verified, which satisfied
	\begin{equation}
		\label{Eq7}
		\frac{{n^*}}{{f_s}}=\frac{\hat{N}}{\hat{B}}.
	\end{equation}
	
	The estimation results for OFDM signals with different \(N\) and \(B\) are also simulated as shown in Fig. \ref{Fig_4}. It can be seen that there exists a certain estimation error under low SNR situation, and \(\hat{N}\) will be more accurate when channel conditions become better. Furthermore, in order to enable efficient modulation and demodulation of OFDM signals based on fast Fourier transform (FFT) and inverse FFT, \(N\) is typically selected from a value set of powers of 2. Therefore, even in the presence of estimation errors under low SNR situation, the accurate estimation of \(N\) remains achievable within the finite set of possible values.
	
	\begin{figure}[!t]
		\centering
		\includegraphics[width=2.5in]{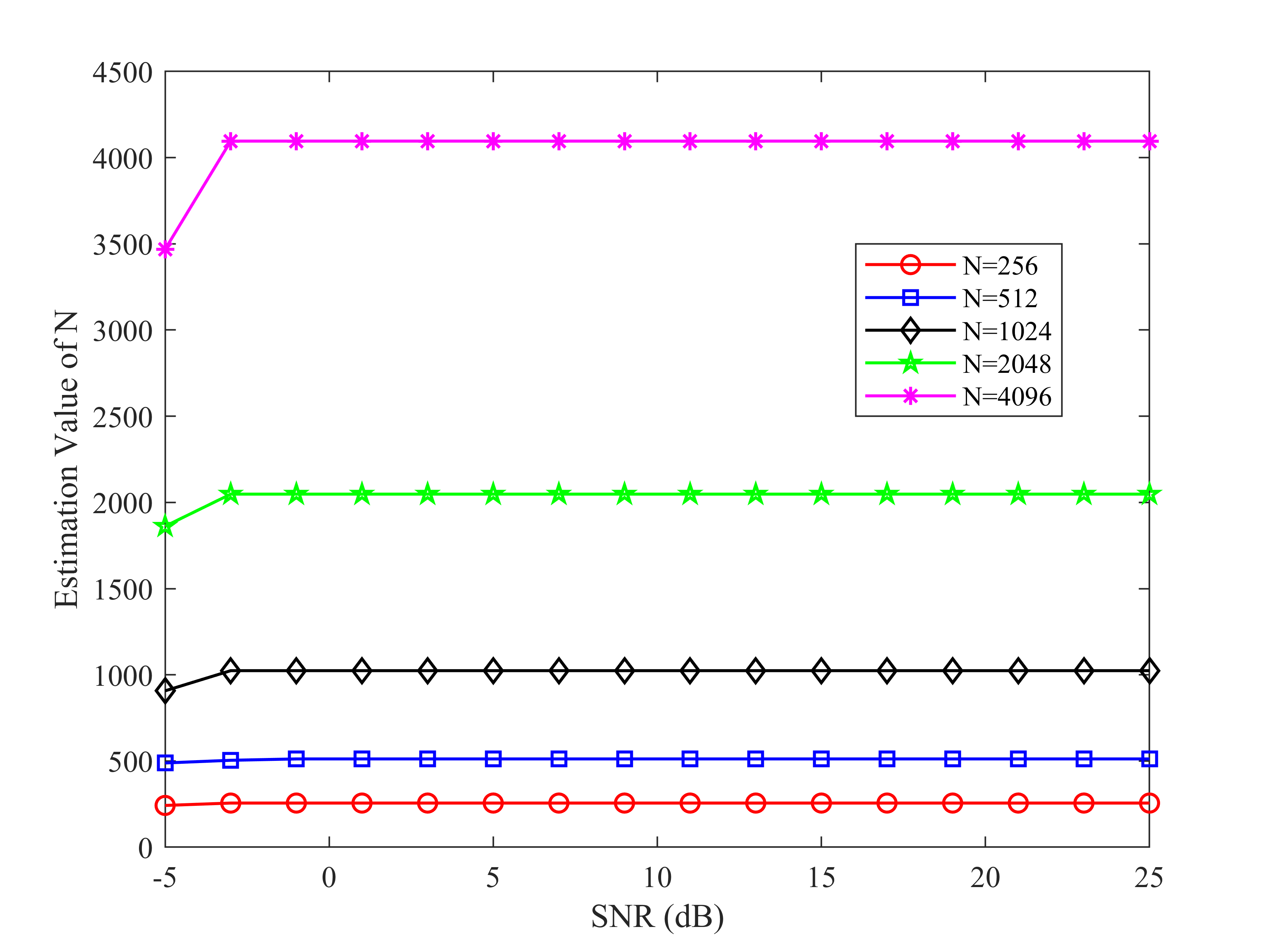}
		\caption{The estimation error of \(\tiny{\hat{N}}\) versus SNR for \(\tiny{f_s=100}\) MHz.}
		\label{Fig_4}
	\end{figure}
	
	\item{\textit{ZC Sequences}}:
	ZC sequences are complex exponential sequences with a constant envelope and remarkable correlation properties, which are widely used in wireless communication for preamble synchronization, channel estimation, and the detection of random access request. ZC sequences can be generated by
	\begin{equation}
		\label{Eq8}
		{z_r}(v)={\text{e}^{-\text{j}\frac{\pi rv(v+1)}{V}}},v=0,1,\ldots ,V-1,
	\end{equation}
	where \(r\) and \(V\) denote the root index and sequence length, respectively, and \(V\) is typically equal to \(N_u + 1\) as regulated. Considering the fact that ZC sequences with different \(r\) exhibit weak cross-correlation characteristic, while the auto-correlation results have a amplitude peak at index \(V\) for zero latency, the \(r\) of the two ZC sequences used in the broadcast frames can be determined. 
\end{enumerate}

\section{Broadcast Frame Detection Algorithm}
\label{sec3}

Various dim and small target enhancement methods in computer vision, such as image reconstruction and multi-dimensional targets separation, can be leveraged to improve the prominence of targets or explore the differences between targets and background noise. However, as a special class of dim and small targets, it should not overlook the significant potential gain in improving drone broadcast frame detection accuracy from the perspective of communication protocols.

\begin{figure}[!t]
	\centering
	\includegraphics[width=2.5in]{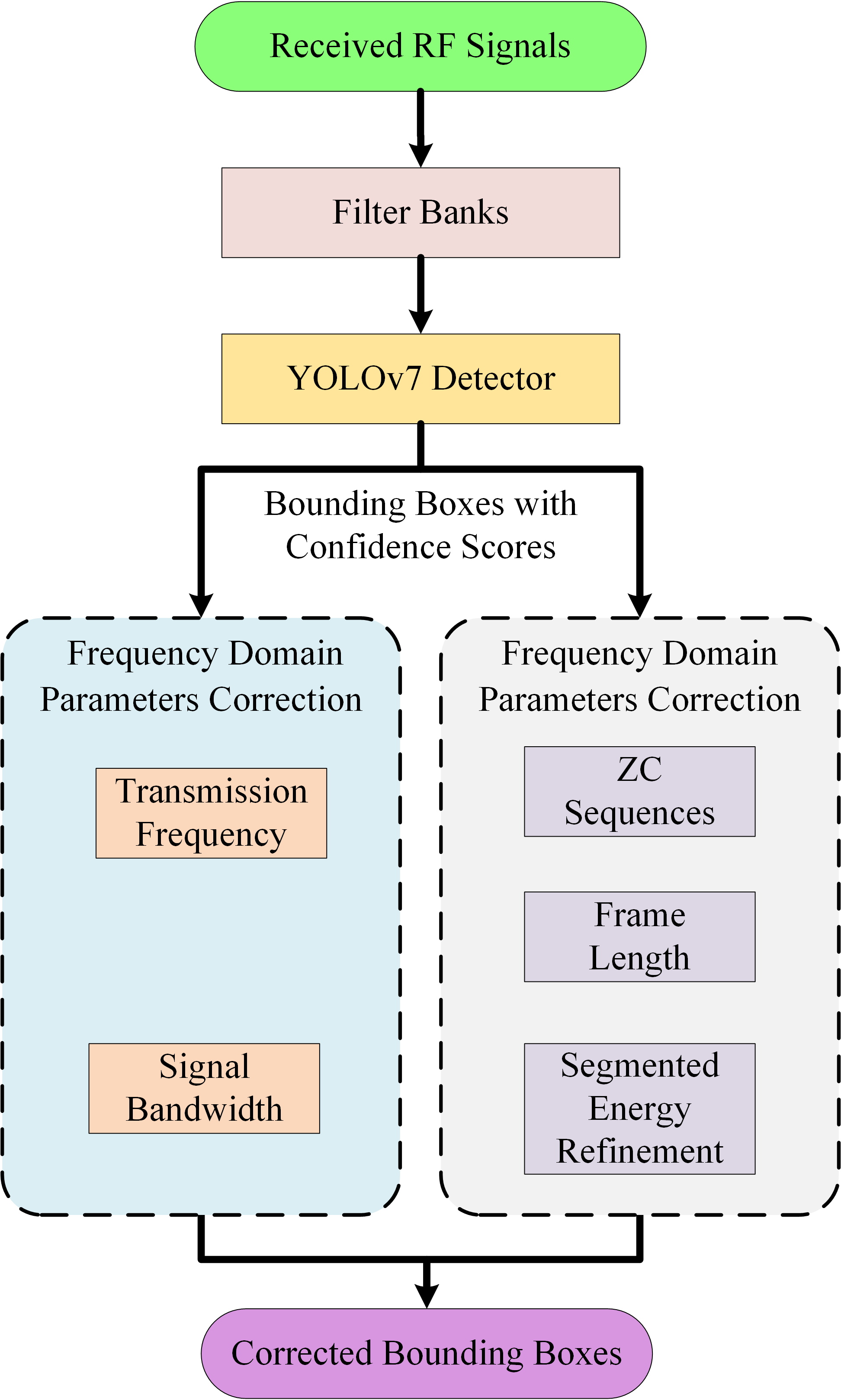}
	\caption{A sketch of the proposed broadcast frame detection algorithm.}
	\label{Fig_5}
\end{figure}

Different from model-level decoupling, the proposed algorithm adopts YOLOv7 as the backbone network and performs data-level decoupling before and after the YOLOv7 detector. Specifically, before being converted into TFIs and sending to the detector, the received RF signals are filtered based on the prior knowledge of transmission frequency and signal bandwidth. After the detector generates the time-frequency parameters and confidence scores of the bounding boxes, all frequency domain parameters are corrected based on the prior knowledge of signal bandwidth and transmission frequency. For bounding boxes with low confidence scores, the time domain parameters are corrected using the cross-correlation results of ZC sequences and the prior knowledge of frame signal length. Additionally, a segmented energy refinement method is applied to mitigate the time domain parameters deviation caused by interference signals with high energy strength, which can further correct the time domain parameters. Fig. \ref{Fig_5} illustrates the architecture of the proposed algorithm.

\subsection{Frequency Domain Parameters Correction}

According to the analysis of the communication protocol, the transmission frequency of drone broadcast frames will be selected from a fixed frequency set, and the signal bandwidth remains constant, which enables the feasibility of frequency domain parameter correction. Specifically, the transmission frequency set \(\mathcal{F}\) and signal bandwidth \(B\) are statistically determined as prior knowledge. On the one hand, \(\mathcal{F}\) and \(B\) are utilized to design filter banks before TFI generation, which can partly eliminate the impact of interference signals and noise from other frequencies on the YOLOv7 detector. On the other hand, \(\mathcal{F}\) and \(B\) are applied to correct the frequency domain parameters of the bounding boxes generated by the detector, which can ensure that the frequency range determined by the frequency domain parameters does not contain other unexpected frequencies.

Let \(f_1 \sim f_4\) and \(f_5 \sim f_8\) denote the transmission frequencies regulated in 2.4 GHz and 5.8 GHz bands, respectively, the transmission frequency bands can be expressed as
\begin{equation}
	\label{Eq9}
	{{F}_{i}}=[{{f}_{i}}-\frac{B}{2},{{f}_{i}}+\frac{B}{2}]\in \mathcal{F}\mathcal{B},1\le i\le 8.
\end{equation}

\begin{figure*}[!t]
	\centering
	\subfloat[]{\includegraphics[width=2.3in]{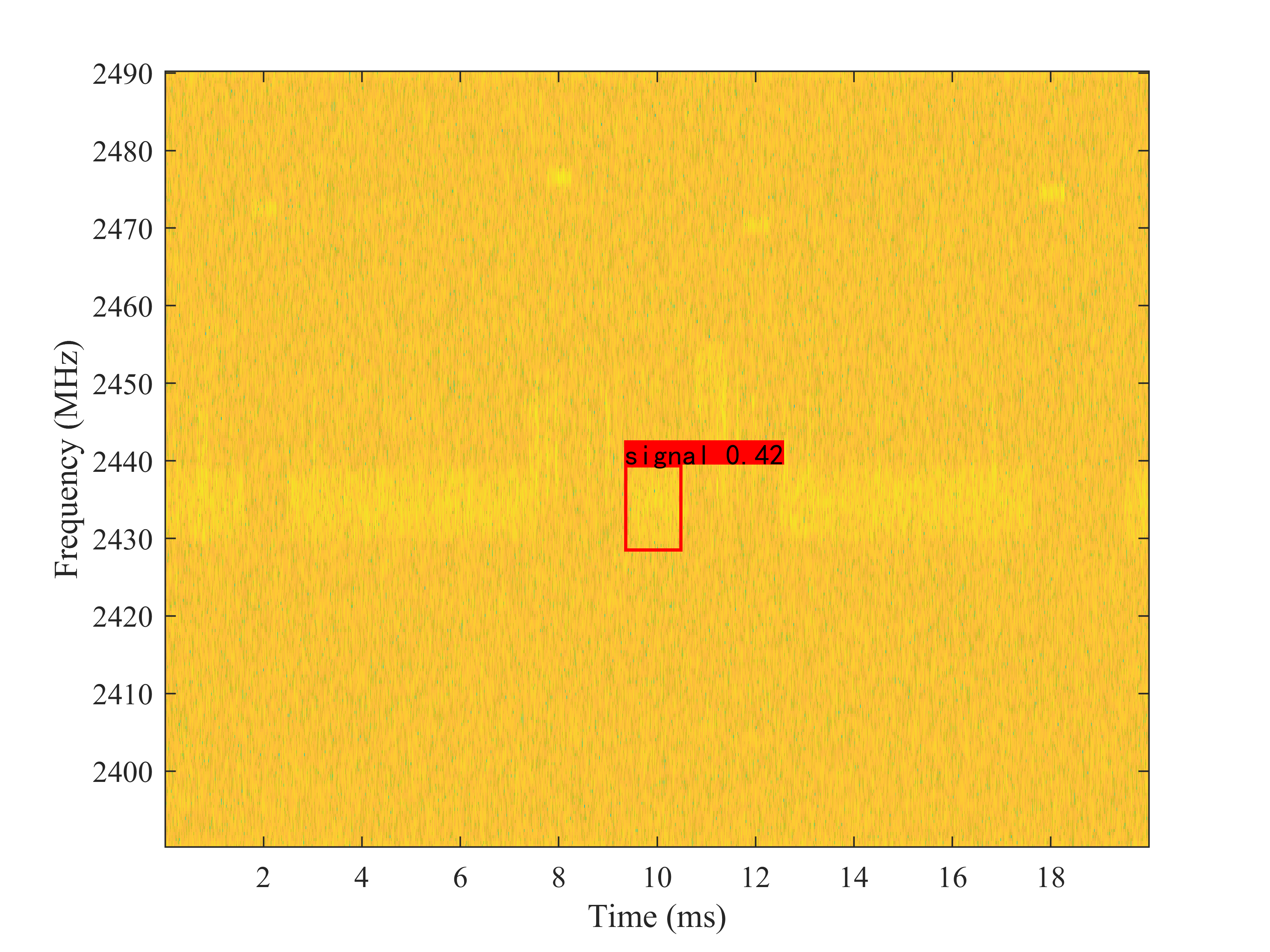}
		\label{67_first}}
	\hfil
	\subfloat[]{\includegraphics[width=2.3in]{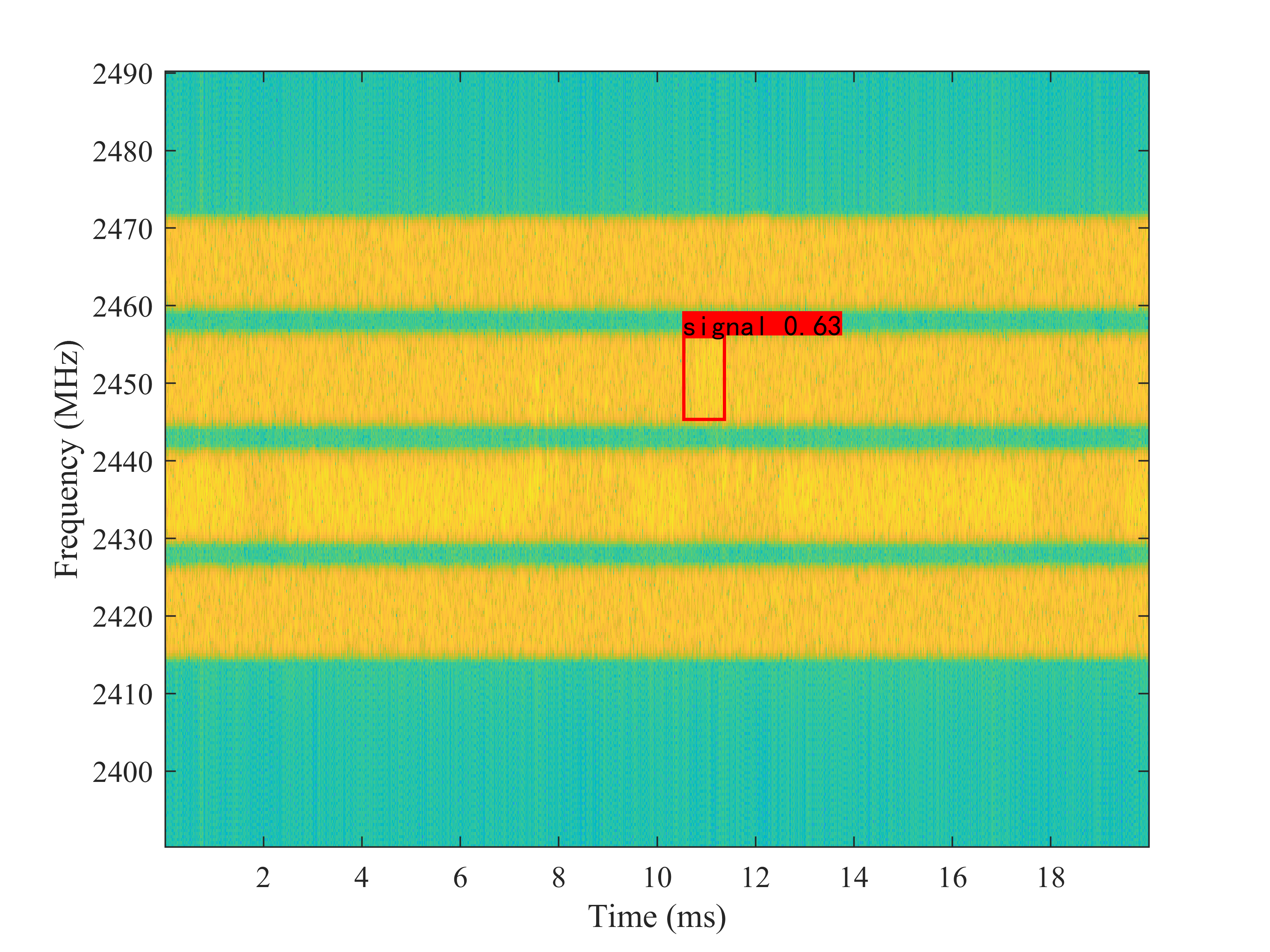}
		\label{67_second}}
	\hfil
	\caption{Bouding box with confidence score of T0011D00S00L4 at SNR = -7 dB. (a) Without filter banks. (b) With filter banks.}
	\label{Fig_6}
\end{figure*}

It is worth noting that the frequency distribution of \(f_1 \sim f_4\) in the 2.4 GHz band and \(f_5 \sim f_8\) in the 5.8 GHz band follows a certain pattern, which can be leveraged to design two sets of band-pass and band-stop filter banks to filter out signals and noise on other frequencies. For different \(f_s\) in DroneRFb-DIR, filters need to be redesigned, so a total of three sets of filter banks are designed. In practical application scenarios, due to the fixed sampling rate of the hardware platform, broadcast frame detection can be achieved by loading pre-designed filter banks without increasing computational costs. As shown in Fig. \ref{67_first}, due to the presence of interference signals, the bounding box shifts toward the downlink OFDM video transmission signals with a lower confidence score. For RF signals that have passed through fiter banks, the bounding box is more accurately positioned on the drone broadcast frame with a higher confidence, as illustrate in Fig. \ref{67_second}.

However, the frequency domain parameters of the bounding boxes generated by the detector for the filtered signals are not entirely accurate. Let \(x_{\min}\), \(x_{\max}\), \(y_{\min}\), and \(y_{\max}\) denote the time domain and frequency domain parameters of a bounding box, respectively, the detected frequency band and bandwidth can be obtained as
\begin{equation}
	\label{Eq10}
	F_d=[y_{\min},y_{\max}], B_d=y_{\max} - y_{\min}.
\end{equation}

\begin{figure*}[!t]
	\centering
	\subfloat[]{\includegraphics[width=2.3in]{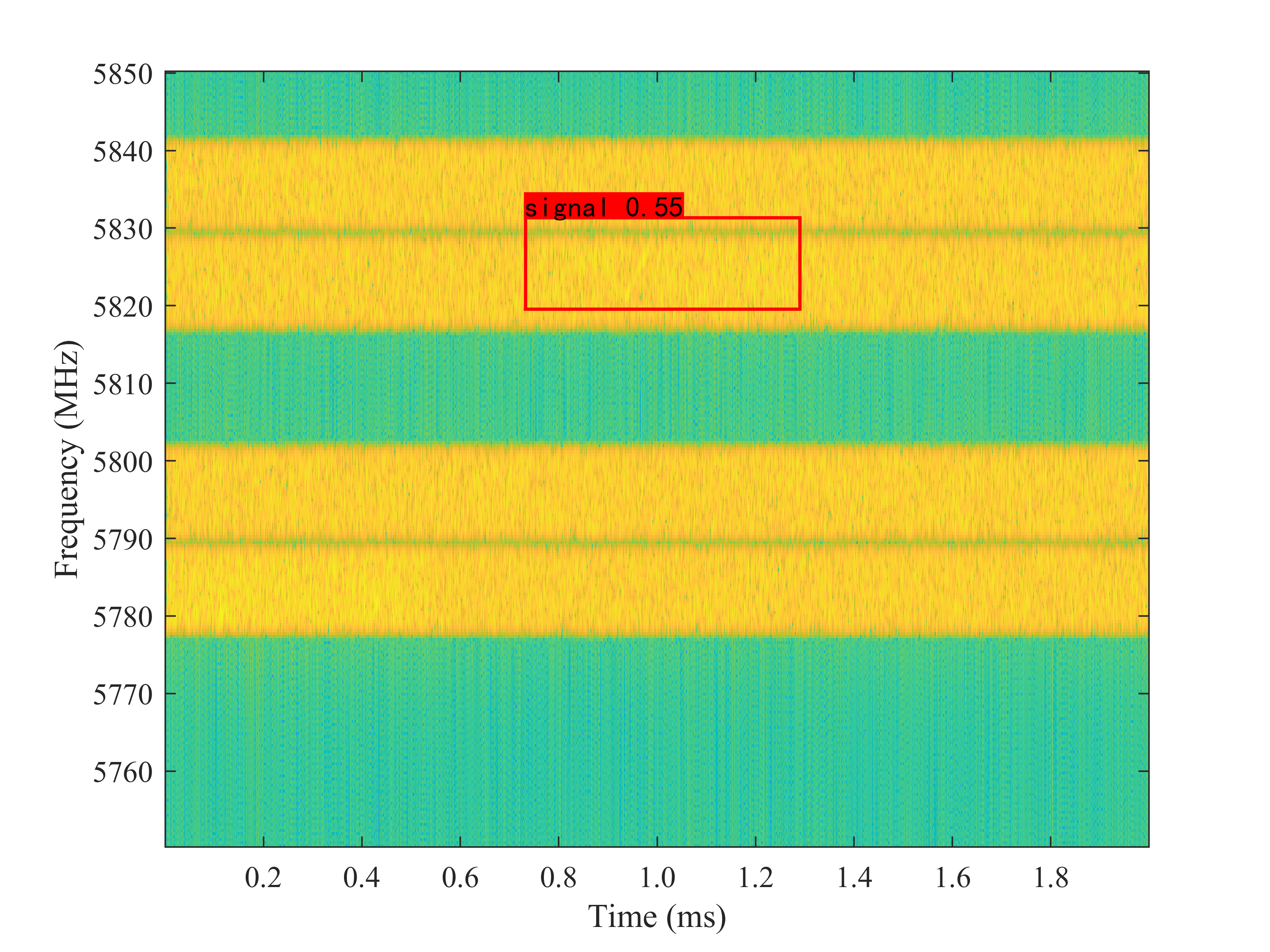}
		\label{89_first}}
	\hfil
	\subfloat[]{\includegraphics[width=2.3in]{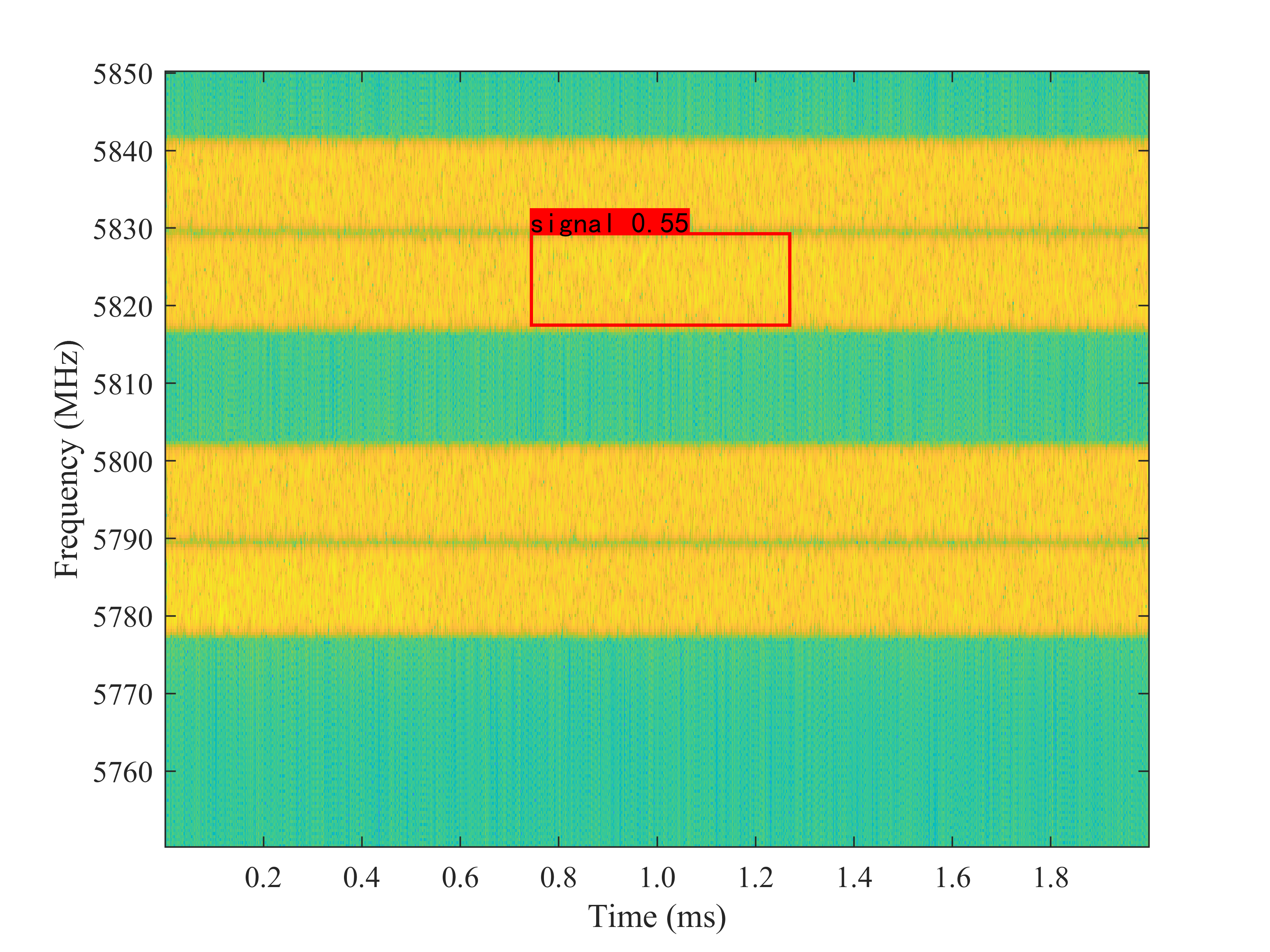}
		\label{89_second}}
	\hfil
	\caption{Bouding box of T0101D00S00L1 at SNR = -7 dB. (a) Original Bounding Box. (b) With frequency domain parameters correction.}
	\label{Fig_7}
\end{figure*}

\begin{figure*}[!t]
	\centering
	\subfloat[]{\includegraphics[width=2.3in]{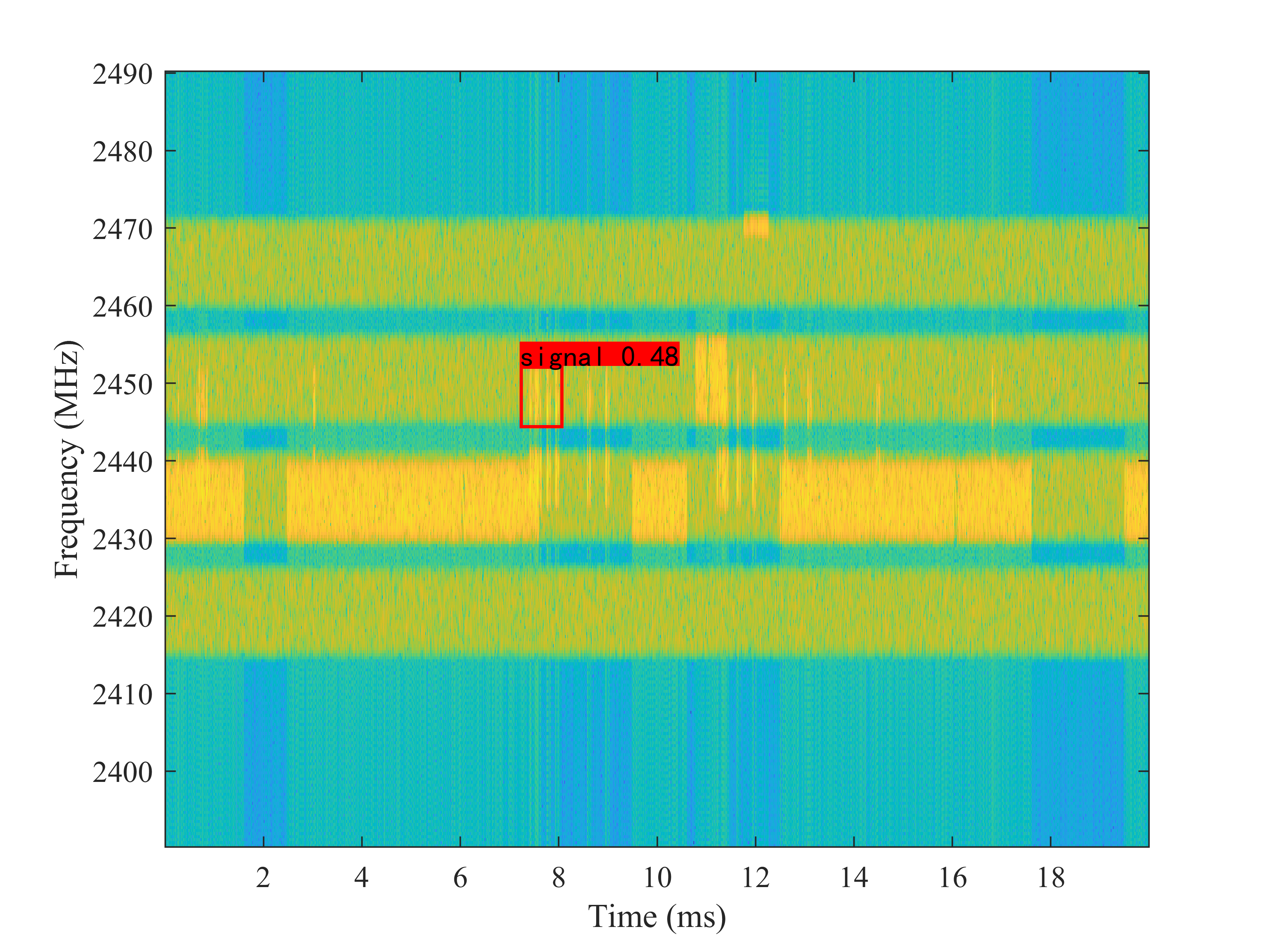}%
		\label{101112_first}}
	\hfil
	\subfloat[]{\includegraphics[width=2.3in]{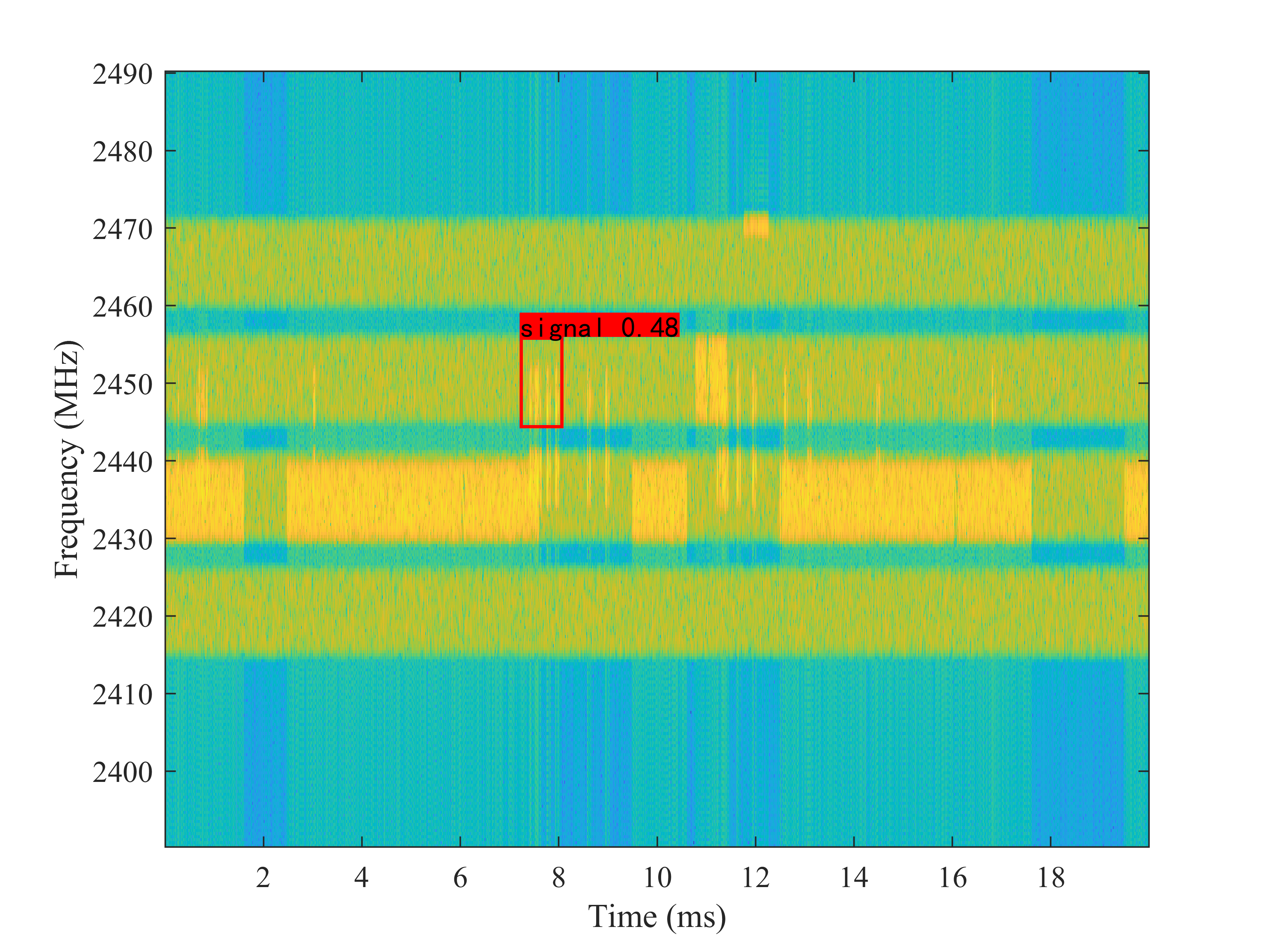}%
		\label{101112_second}}
	\hfil
	\subfloat[]{\includegraphics[width=2.3in]{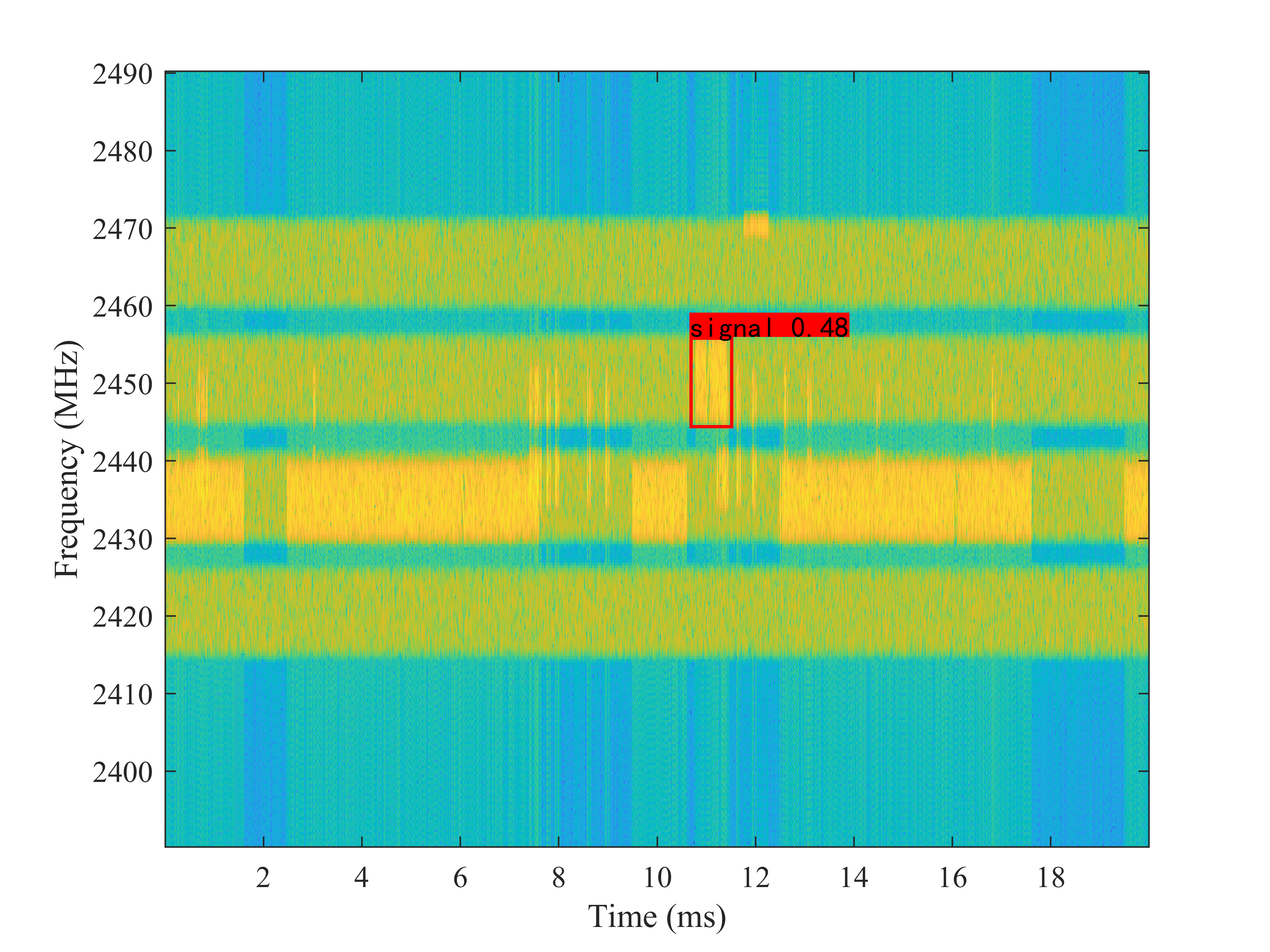}%
		\label{101112_third}}
	\caption{Bouding box of T0011D00S00L4 at SNR = 15 dB. (a) Original Bounding Box. (b) With frequency domain correction. (c) With time domain correction.}
	\label{Fig_8}
\end{figure*}

There are two types of frequency domain parameter deviation from ground truth, one situation is as shown in Fig. \ref{89_first}, where \({F_d}\notin \mathcal{F}\mathcal{B}\) due to the presence of noise, and it is evident that there exists \(y_{\min}>{f_i}-\frac{B}{2}\) and \(y_{\max}>{f_i}+\frac{B}{2}\). Then \(y_{\min}\) and \(y_{\max}\) will be corrected by
\begin{equation}
	\label{Eq11}
	\left\{
	\begin{array}{cc}
		\hat{y}_{\max}=\frac{y_{\max}+y_{\min}}{2}+\frac{B}{2}, & \frac{y_{\max}+y_{\min}}{2}\le f_i\\
		\hat{y}_{\max}=f_i+\frac{B}{2}, & \frac{y_{\max}+y_{\min}}{2}> f_i,
	\end{array}
	\right.
\end{equation}
where \(\hat{y}_{\max}-\hat{y}_{\min} = B\) is established to correct \(y_{\min}\) and \(y_{\max}\) under the situation that \(F_d \ne B\) due to the existence of interference signals, as shown in Fig. \ref{101112_first}.

Assuming the length of filter is \(L_f\), the computational complexity of frequency domain parameters correction is \(\mathcal{O}( 2LL_f )\) with disregarding the complexity of the YOLOv7 detector.

\subsection{Time Domain Parameters Correction}

Due to the presence of noise and interference signals, the time domain parameters of the bounding boxes may deviate from the ground truth. The detected duration of the drone broadcast frame is \(T_d=x_{\max} - x_{\min}\), while the regulated duration can be computed by \(T=\frac{M(N+N_{\text{cp}})}{B}\), where \(M\) denotes the number of OFDM symbols in one broadcast frame. It is worth noting that, according to the prior knowledge of the communication protocol, \(M\) and \(N_{\text{cp}}\) will vary based on the drone type and the OFDM symbol index.

For the case that the bounding box is generated at the non-broadcast frame position, as shown in Fig. \ref{101112_second}, the deviation of \(x_{\max}\) and \(x_{\min}\) will be corrected by ZC sequences adopted in the broadcast frames. Specifically, the cross-correlation operation between the filter RF signals and locally generated ZC sequences can be expressed as
\begin{equation}
	\label{Eq12}
	\gamma(m)=\left| \sum\limits_{k=1}^{V}{z^*_r(k)\hat{x}(k+m)} \right|,m=0,1,\ldots ,L-V-1,
\end{equation}
where \(\hat{x}(m)\) denotes the RF signals \(x(n)\) after passing through the filter banks. Let \(m^*\) denote the index of the peak of \(\gamma(m)\), which is also the beginning of the OFDM symbol contains ZC sequences, and the corrected time domain parameters of the detected broadcast frame can be given by
\begin{equation}
	\label{Eq13}
	\left\{
	\begin{array}{c}
		\hat{x}_{\min}=m^*-\lambda(N+N_{\text{normal}})-N_{\text{extend}}\\
		\hat{y}_{\min}=\hat{x}_{\min}+T,
	\end{array}
	\right.
\end{equation}
where \(N_{\text{normal}}\) and \(N_{\text{extend}}\) denote the different length of \(N_{\text{cp}}\), and \(\lambda\) is the parameter obtained by analyzing \(M\) of different types of drone's broadcast frames.

\begin{figure*}[!t]
	\centering
	\subfloat[]{\includegraphics[width=2.3in]{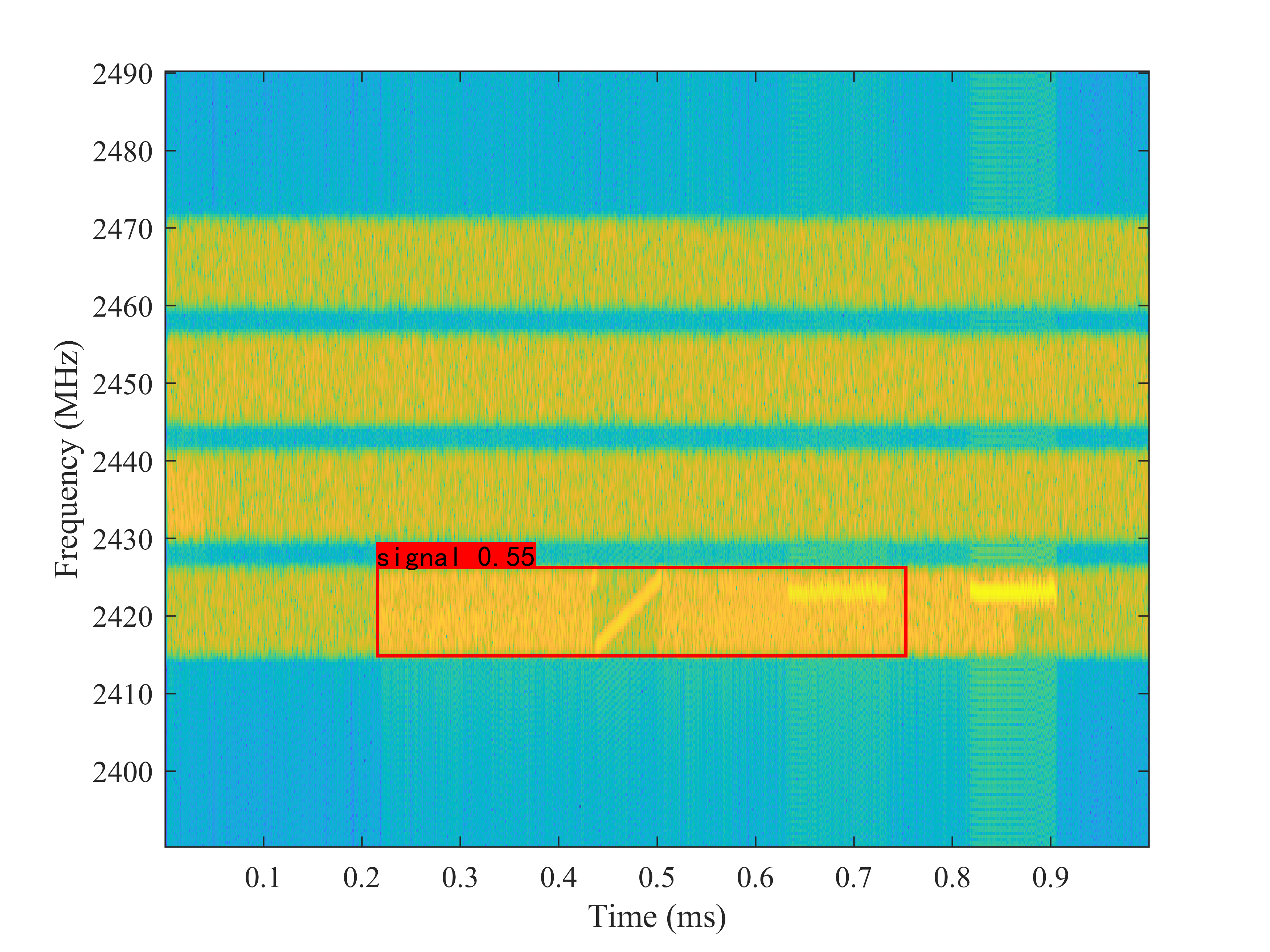}
		\label{1314_first}}
	\hfil
	\subfloat[]{\includegraphics[width=2.3in]{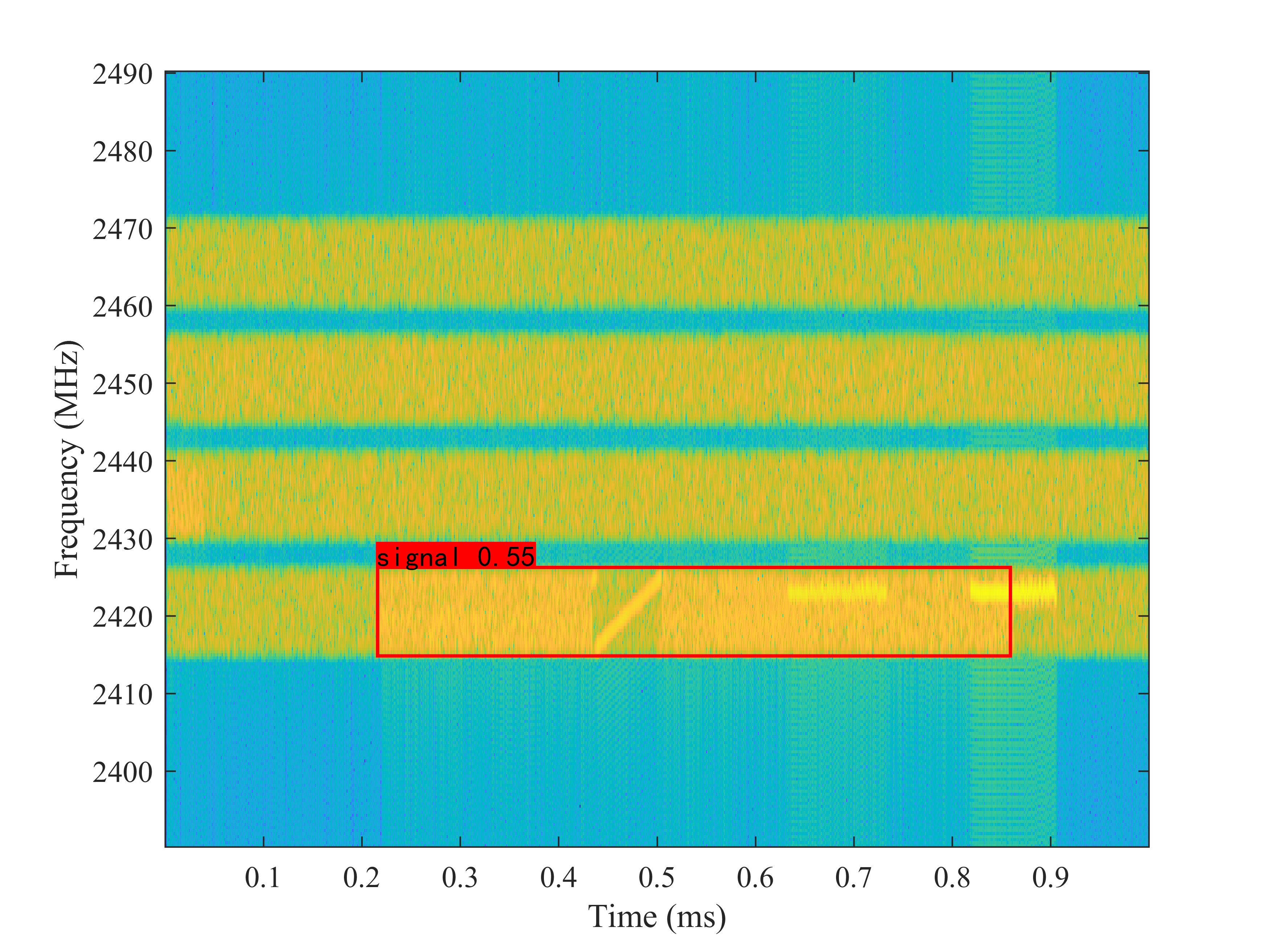}
		\label{1314_second}}
	\hfil
	\caption{Bouding box of T0011D00S00L0 at SNR = 15 dB. (a) Original Bounding Box. (b) With time domain parameters correction.}
	\label{Fig_9}
\end{figure*}

\begin{figure*}[!t]
	\centering
	\subfloat[]{\includegraphics[width=2.3in]{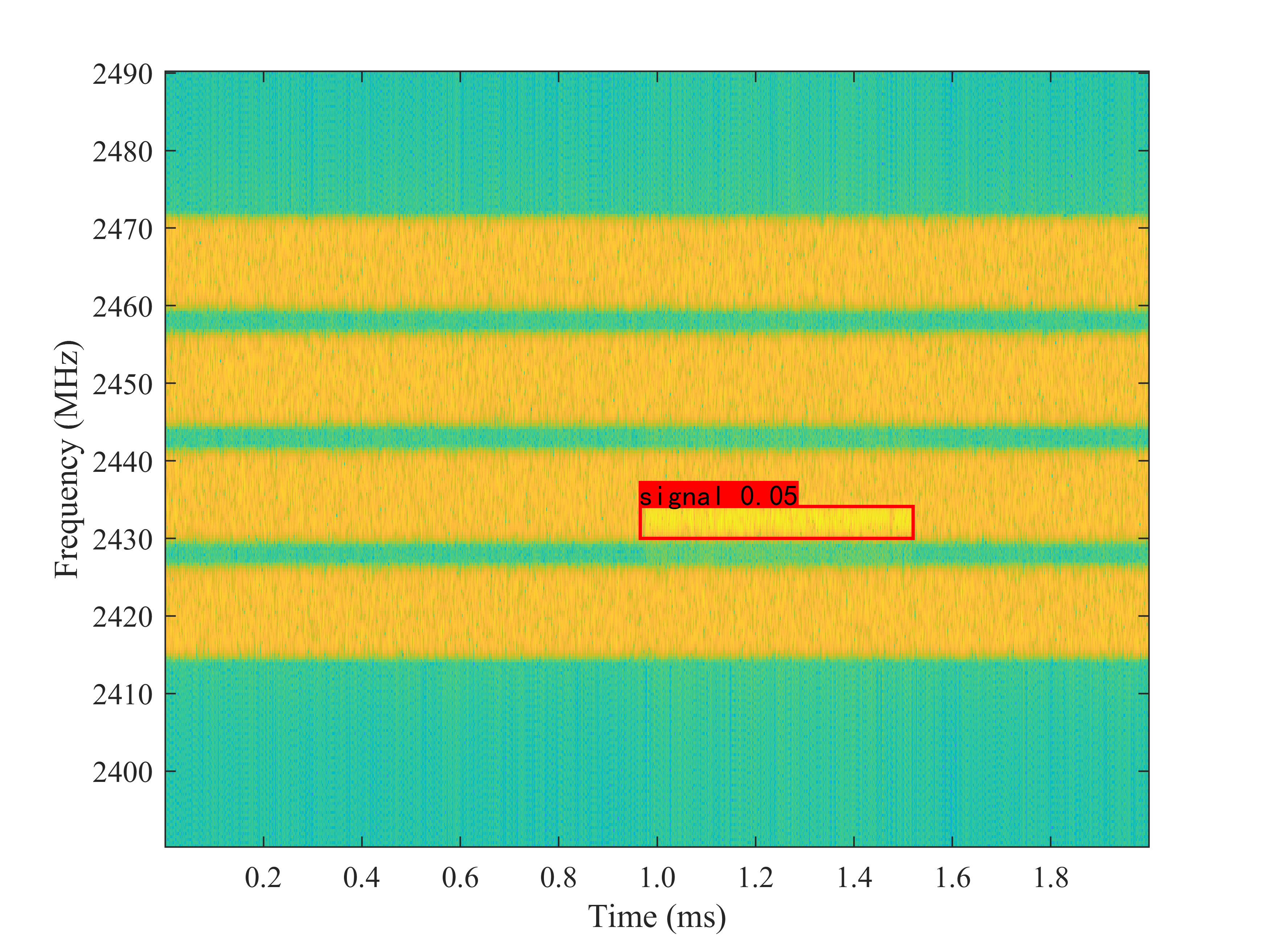}%
		\label{151617_first}}
	\hfil
	\subfloat[]{\includegraphics[width=2.3in]{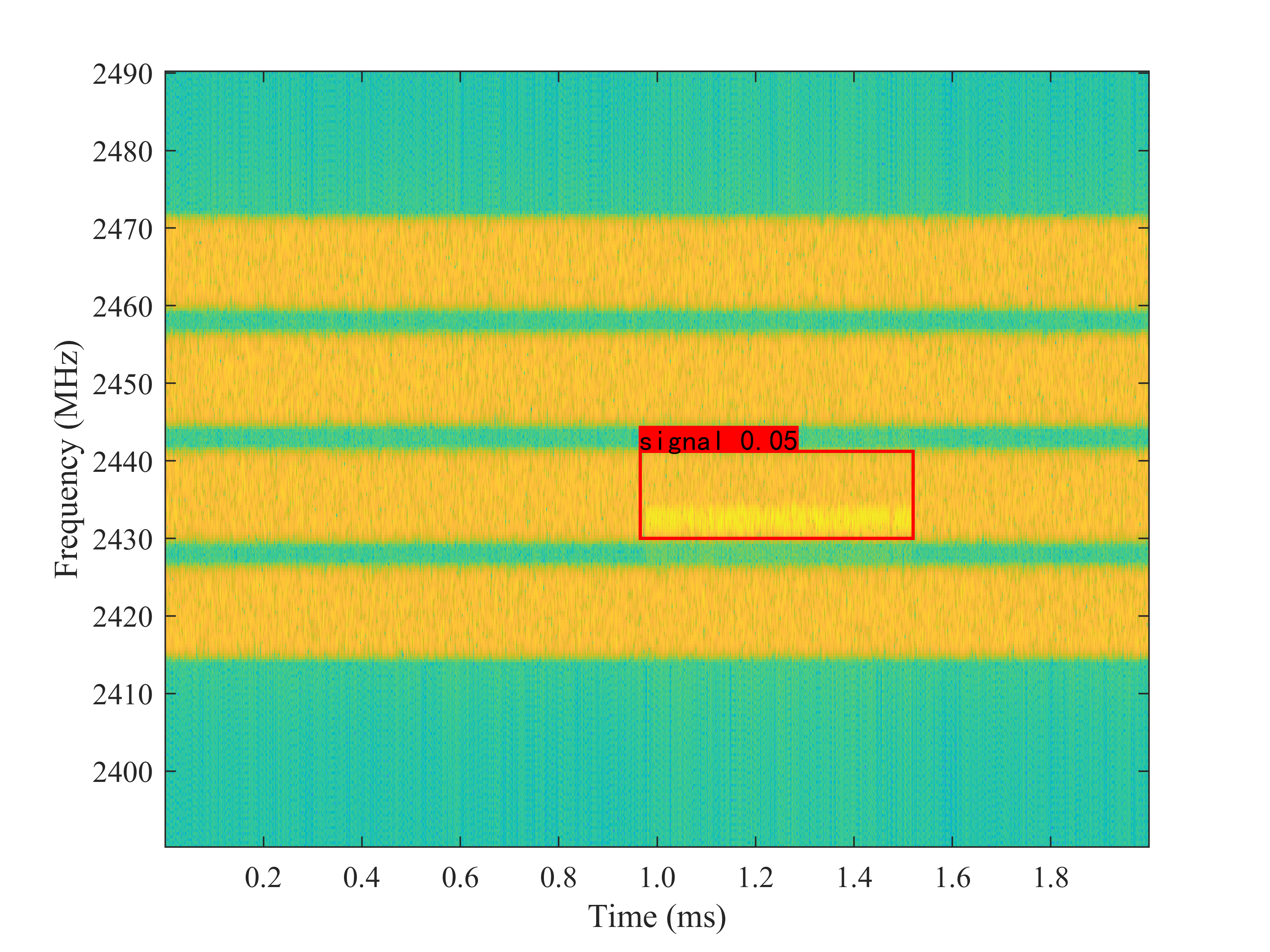}%
		\label{151617_second}}
	\hfil
	\subfloat[]{\includegraphics[width=2.3in]{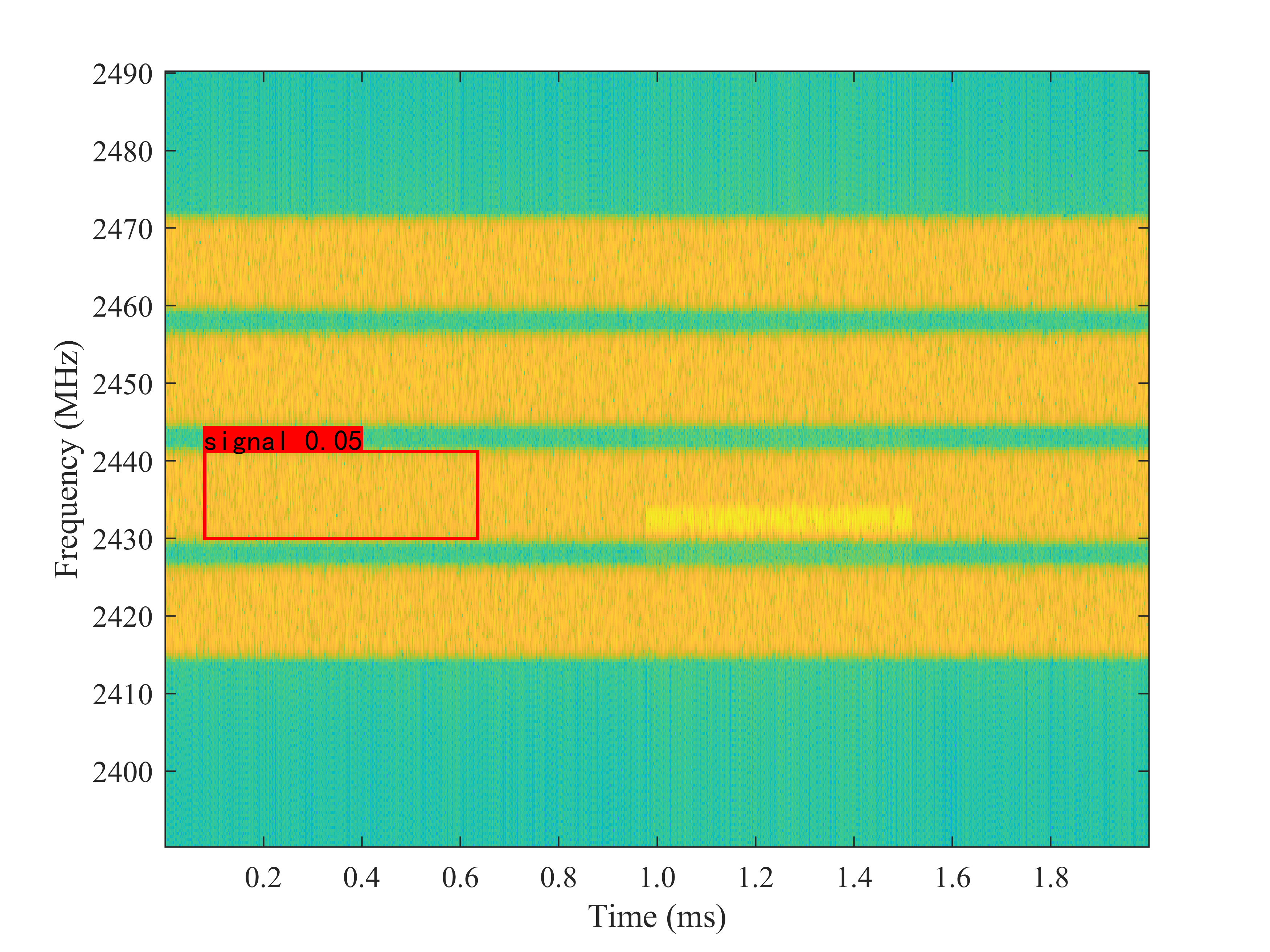}%
		\label{151617_third}}
	\caption{Bouding box T0001D00S00L1 at SNR = -9 dB. (a) Original Bounding Box. (b) With frequency domain correction. (c) With time domain correction.}
	\label{Fig_10}
\end{figure*}

\begin{figure*}[!t]
	\centering
	\subfloat[]{\includegraphics[width=2.3in]{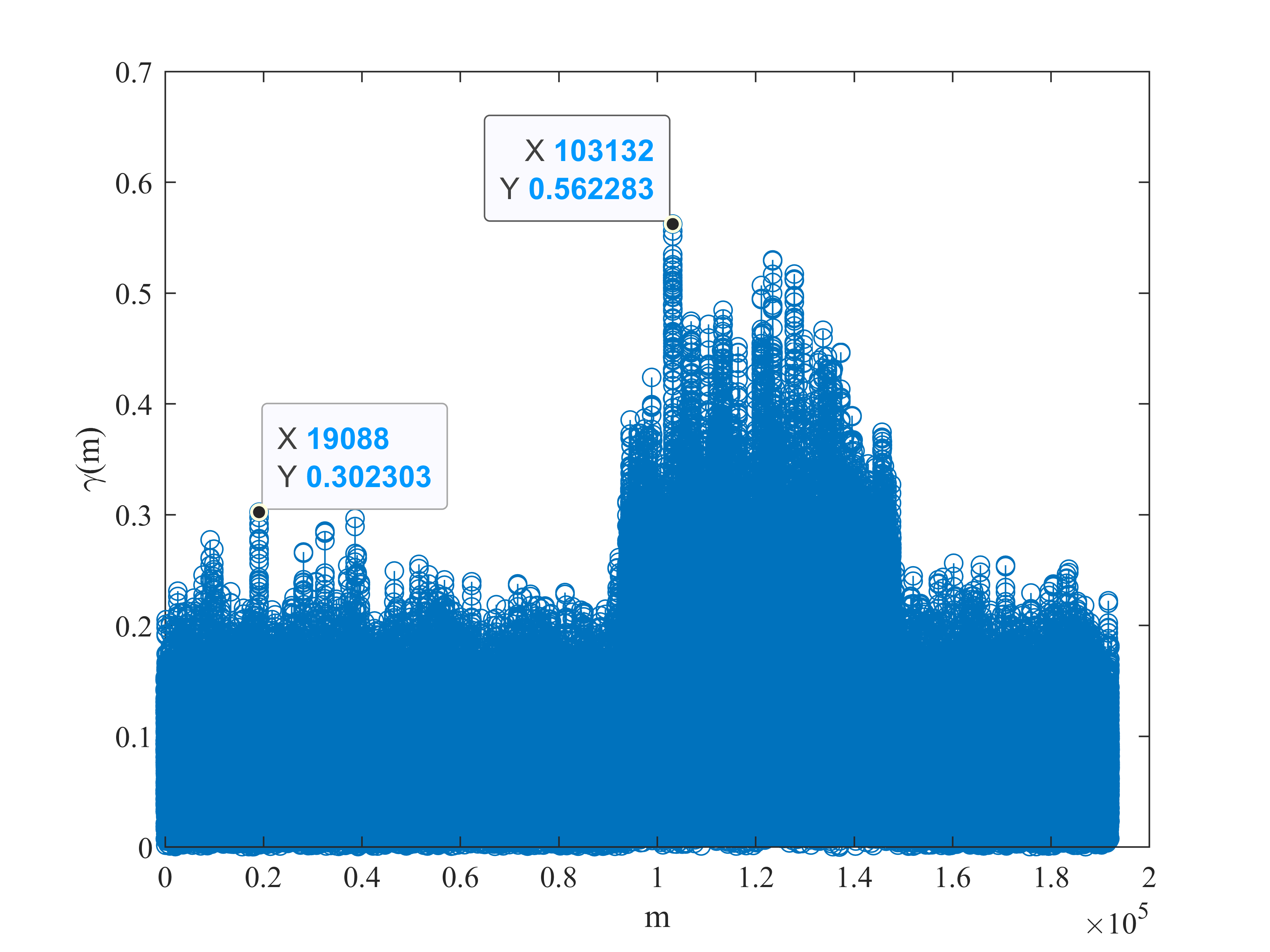}%
		\label{181920_first}}
	\hfil
	\subfloat[]{\includegraphics[width=2.3in]{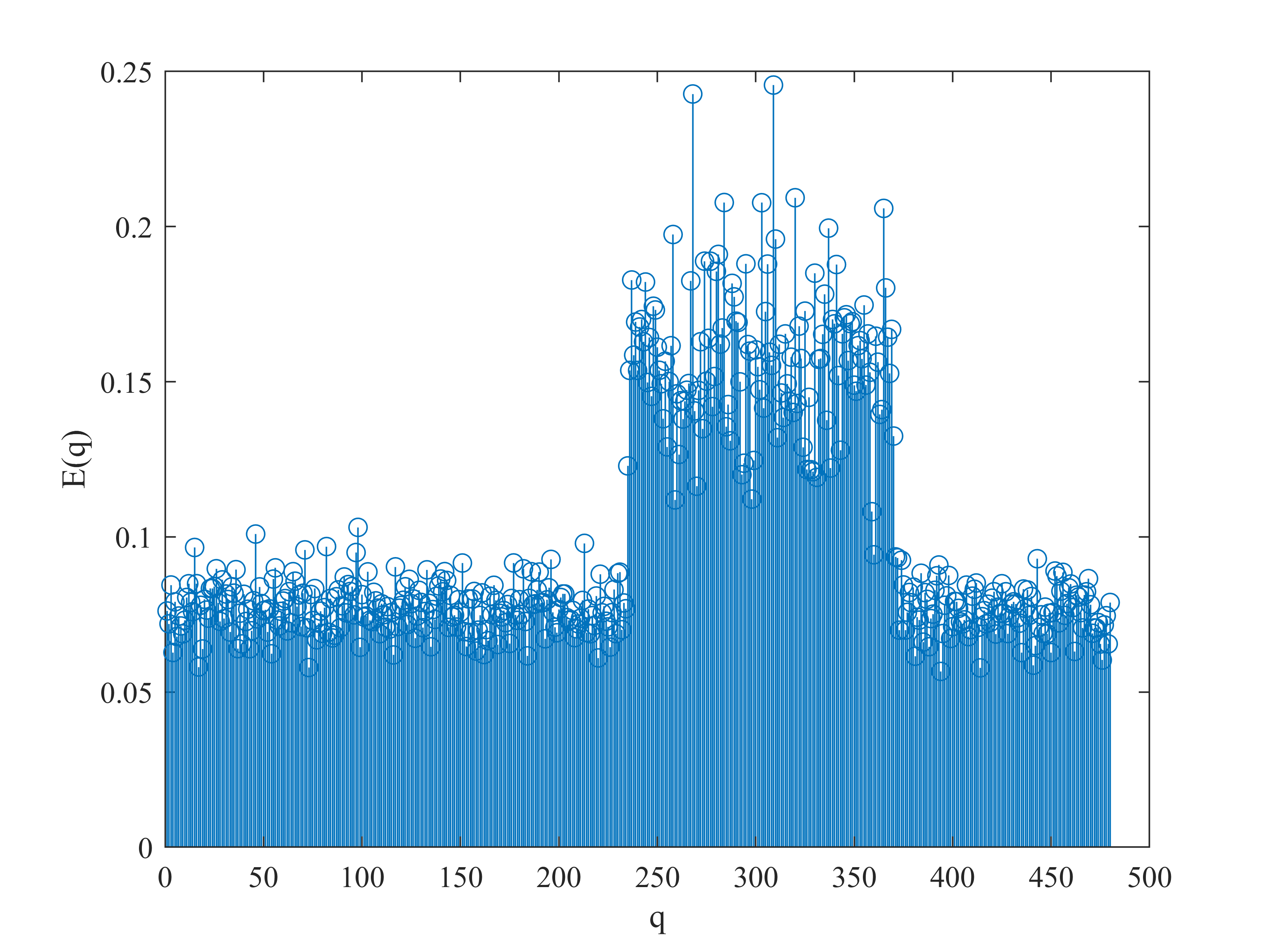}%
		\label{181920_second}}
	\hfil
	\subfloat[]{\includegraphics[width=2.3in]{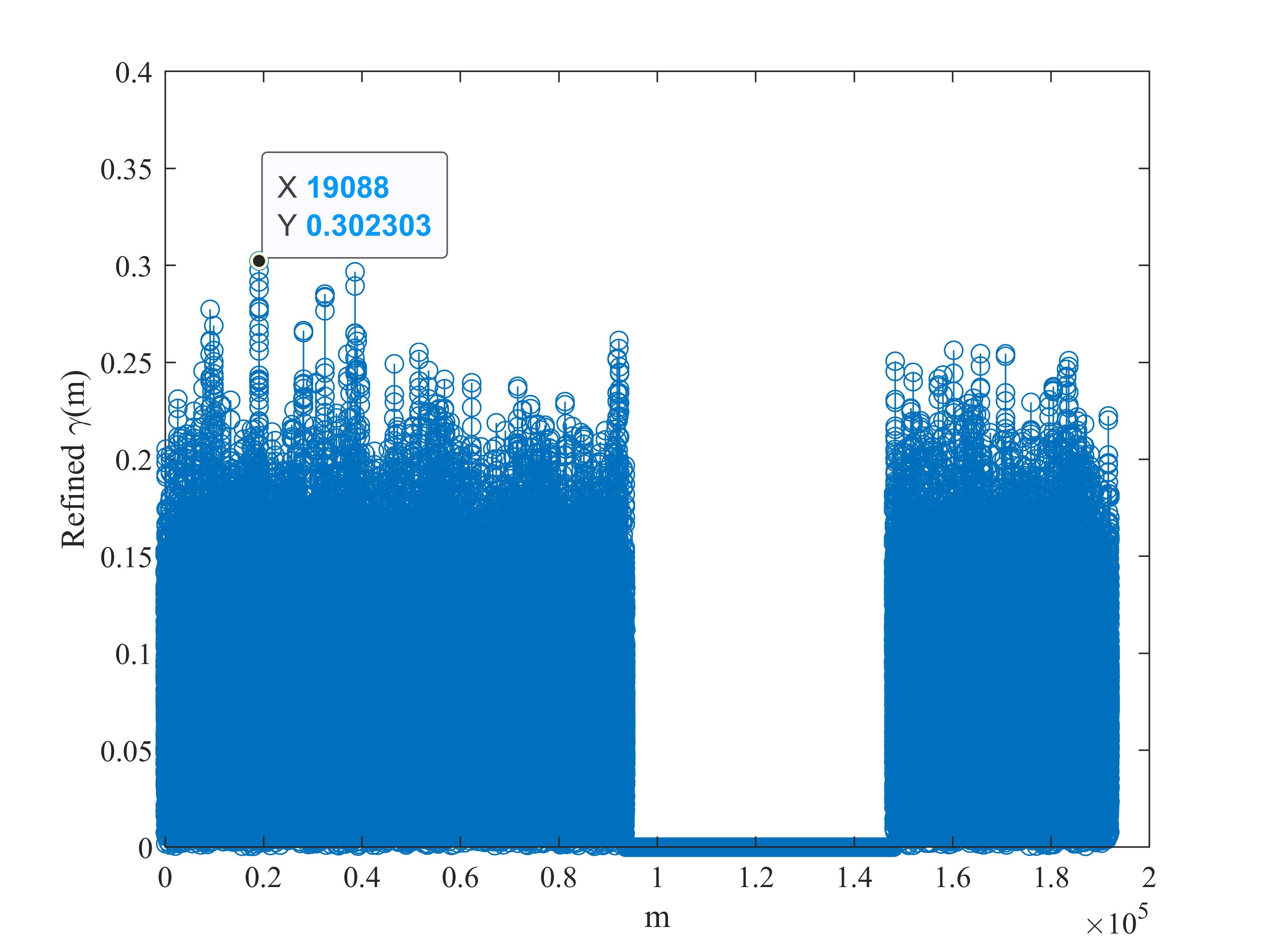}%
		\label{181920_third}}
	\caption{Time domain parameters correction of T0001D00S00L1 at SNR = -9 dB. (a) \(\tiny{\gamma(m)}\). (b) \(\tiny{E(q)}\). (c) \(\tiny{\hat{\gamma}(m)}\).}
	\label{Fig_11}
\end{figure*}

For time domain parameter deviation caused by the presence of interference signals, as shown in Fig. \ref{1314_first}, the correction of \(x_{\min}\) and \(x_{\max}\) can be performed directly based on \textcolor{blue}{(\ref{Eq13})}. However, under low SNR situations, bounding boxes may be generated at the time-frequency positions of interference signals with high energy strength, as shown in Fig. \ref{151617_first}. Although \(y_{\min}\) and \(y_{\max}\) can be corrected by \textcolor{blue}{(\ref{Eq11})}, the correction of \(x_{\min}\) and \(x_{\max}\) based on \textcolor{blue}{(\ref{Eq13})} merely will result in the wrong \(m^*\) corresponding to the interference signals, as shown in  Fig. \ref{181920_first}. Thus, a segmented energy refinement method is applied to mitigate the time domain parameters deviation caused by the interference signals with high energy strength. \(\gamma(m)\) will be divided into \(Q\) non-overlapping sequences with length \(P\), and the mean energy of each segment can be calculated by
\begin{equation}
	\label{Eq14}
	E(q) = \frac{1}{P}\sum\limits_{p=1}^{P}{\gamma[(q-1)P+p]},q=1,2,\ldots ,Q, 
\end{equation}
where \(PQ=L-V\).

It can be observed that for ZC sequences with short duration, the cross-correlation peak values within the corresponding segments have small gaps relative to those of non-ZC sequence segment. In contrast, for interference signals with longer duration, which have high energy strength, the peak values within the corresponding segments have bigger gaps than those of other segments. Let \(E_Q\) denote the average value of \(E(q)\), the index set \(I\) of the interference signals can be expressed as
\begin{equation}
	\label{Eq15}
 	\left\{
 	\begin{array}{cc}
 		q \in I, & E(q)\ge \alpha E_Q\\
 		q \notin I, & E(q)< \alpha E_Q,
 	\end{array}
 	\right.
\end{equation}
where \(\alpha\) denotes the weighting factor to balance the detection accuracy under different SNR. Since the elements in \(I\) corresponding to the interference signals may be non-contiguous due to a large number of low intensity \(\gamma(m)\), consecutive \(q\) will be added when the difference between contiguous \(q \in I\) is less than \(\beta\).

The \(\gamma(m)\) contained within \(q \in I\) will be refined by
\begin{equation}
	\label{Eq16}
	\left\{
	\begin{array}{cc}
		\hat{\gamma}[(q-1)P+(1\!:\!P)] = \gamma[(q-1)P+(1\!:\!P)], & q \notin I\\
		\hat{\gamma}[(q-1)P+(1\!:\!P)] = 0, & q \in I.
	\end{array}
	\right.
\end{equation}

The simulation results of \(\hat{\gamma}(m)\) and the corrected bounding box are shown in Fig. \ref{151617_third} and Fig. \ref{181920_third}, it can be seen that the influence of interference signals with high energy strength on the time domain parameters has been eliminated. Although \(E_Q\) is also the average value of \(\gamma(m)\), the segmented refinement method differs from the direct refinement method based \textcolor{blue}{(\ref{Eq17})}. This is because \textcolor{blue}{(\ref{Eq16})} can better enhance the cross-correlation intensity difference between ZC sequences and the interference signals. Specifically, most \(\gamma(m)\) for \(m\in [0,L-V-1]\) have relatively low intensity, both ZC sequences and interference signals will be identified as outliers and refined, which results in wrong time domain parameters and low detection accuracy of the drone broadcast frames.
\begin{equation}
	\label{Eq17}
	\left\{
	\begin{array}{cc}
		\hat{\gamma}(m) = \gamma(m), & \gamma(m) < \alpha E_Q\\
		\hat{\gamma}(m) = 0, & \gamma(m) \ge \alpha E_Q.
	\end{array}
	\right.
\end{equation}

The computational complexity of time domain parameters correction is \(\mathcal{O}( L(V+2)+Q )\), and the overall computational complexity for the proposed algorithm without including the detector is \(\mathcal{O}( L(2L_f+V+2)+Q )\).

\section{Broadcast Frame Decoding Algorithm}
\label{sec4}
For a broadcast frame transmitted at \(f_i\), the bitstream is obtained after performing time and frequency synchronization, which is then decoded to recover the payload information.
\subsection{Time and Frequency Synchronization}
\begin{enumerate}[leftmargin=0pt, itemindent=2pc, listparindent=\parindent]
	\item{\textit{Frequency Coarse Synchronization}}: The signal frequency is shifted from \(f_i\) to near-zero, then \(\hat{x}(n)\) will pass through a low-pass filter and be resample to obtain \(\dot{x}(m)\).
	\item{\textit{Time Coarse Synchronization}}: The start of OFDM symbols can be determined by the peak of auto-correlated results of CP, which can be expressed as
	\begin{equation}
		\label{Eq18}
		\gamma (m)=\frac{{{\left| \sum\limits_{n=1}^{{N_{\text{cp}}}}{\dot{x}(m+n){{{\dot{x}}}^{*}}(m+n+N)} \right|}^{2}}}{\left[ \sum\limits_{n=1}^{{N_{\text{cp}}}}{\dot{x}(m+n){{{\dot{x}}}^{*}}(m+n)} \right]\left[ \sum\limits_{n=1}^{{N_{\text{cp}}}}{\dot{x}(m+n+N){{{\dot{x}}}^{*}}(m+n+N)} \right]}. 
	\end{equation}
	\item{\textit{Frequency Fine Synchronization}}: The frequency offset \(\varepsilon\) can be estimated by \(-\frac{\theta N}{2\pi h}\), where \(\theta\) denote the phase angle at the peak of cross-correlation results based on CP or ZC sequences, and \(h\) can be either \(N_\text{cp}\) or \(N\). Then we have
	\begin{equation}
		\label{Eq19}
		\ddot{x}(n)=\dot{x}(n)\text{e}^{-\frac{\text{j}2\pi t}{N}\varepsilon},t=1,2,\ldots ,T.
	\end{equation}
	\item{\textit{Time Fine Synchronization}}: The start of symbols will be re-estimated based on \(\ddot{x}(n)\) and \textcolor{blue}{(\ref{Eq18})}. Subsequently, CP will be removed, and OFDM symbols are demodulated and quantized to obtain bitstream.
	
\end{enumerate}
\subsection{Bitstream Decode}
The bitstream is descrambled based on the Gold sequences and subsequently processed by a Turbo decoder to recover the payload information\textcolor{blue}{\cite{ref14}},\textcolor{blue}{\cite{ref15}},\textcolor{blue}{\cite{ref52}}, and the cyclic redundancy check (CRC) value contained in each frame will confirms the correctness of decoded result. In addition, the remaining data can also verify the correctness of decoding results. Since the datasets come from Zhejiang University, the latitude and longitude in the decoding results should match it. The first three symbols of the serial number corresponding to different types of drones are different and fixed, thus the decoding result should also be consistent with the regulations.

\section{Numerical Results}
\label{sec5}
\subsection{Simulation Setup}

RF signals from different types of drone, various flight distances, diverse flight visual environment, and different sampling duration are used as raw samples. Besides, four types of noise are introduced to evaluate the robustness of the proposed algorithm under different SNR conditions. SNR varies from -15 dB to 15 dB in 2 dB intervals, and the samples are split into training, validation, and test sets in a 8:1:1 ratio. The hardware environment consists of an NVIDIA GeForce RTX 4060 Ti GPU and an 11th Gen Intel(R) Core(TM) i7-11700K @ 3.60GHz, with the remaining simulation parameters are provided in Table \ref{tab:table2}.

\begin{table}[!t]
	\caption{Simulation Parameters \label{tab:table2}}
	\centering
	\begin{tabular}{c c}
		\hline
		Parameter & Value \\
		\hline
		Batchsize for U-Net & 16 \\
		Batchsize for YOLOv7 & 8 \\
		Epoch for U-Net & 50 \\
		Epoch for YOLOv7 & 100 \\
		Learning Rate & 0.0001 \\
		\hline
	\end{tabular}
\end{table}

The baseline and the proposed algorithms are given as follows.
\begin{enumerate}[leftmargin=0pt, itemindent=2pc, listparindent=\parindent]
	\item{\textit{YOLOv7 (YOLO)}}: The I/Q samples are converted into TFIs directly, and the YOLOv7 network detect the position of drone broadcast frames without any time and frequency domain parameters correction.
	
	\item{\textit{Reconstruction-Based YOLOv7 (Reconstruction-YOLO)}}: The TFIs consist of broadcast frames, noise, and interference signals will be reconstructed by the pre-trained U-shaped (U-Net) network, and the TFIs with denoising and interference-removal will be detected by the YOLOv7 network without any time and frequency domain parameters correction.
	
	\item{\textit{Frequency Domain Parameteres Correction on Model-Level-Based YOLOv7 (FM-YOLO)}}: TFIs are detected by the YOLOv7 network directly, and the prior knowledge of transmission frequency and signal band will be utilized to correct the frequency domain parameters.
	
	\item{\textit{Frequency Domain Parameteres Correction on Data-Level-Based YOLOv7 (FD-YOLO)}}: The I/Q samples will pass through filter banks and be converted into TFIs, and the frequency domain parameters of the bounding boxes generated by the YOLOv7 detector will be corrected.
	
	\item{\textit{Time and Frequency Domain Parameteres Correctionon Model-Level-Based YOLOv7 (TFM-YOLO)}}: TFIs converted from I/Q samples without filter banks are detected by the YOLOv7 network directly, then both the time and frequency domain parameters will be corrected. 
	
	\item{\textit{Time and Frequency Domain Parameteres Correctionon Data-Level-Based YOLOv7 (TFD-YOLO)}}: The filter RF signals are converted into TFIs, then both the time and frequency domain parameters of the bounding boxes generated by the YOLOv7 detector will be corrected.
	
	The drone broadcast frames detection will be evaluated by the metrics as IoU, precision, recall, and weighted evaluation metric, which can be computed by
	\begin{equation}
		\label{Eq20}
		\text{IoU}=\frac{\left| A\bigcap B \right|}{\left| A\bigcup B \right|}, 
	\end{equation}
	\begin{equation}
		\label{Eq21}
		\text{Precision}=\frac{T\!P}{T\!P+F\!P}, 
	\end{equation}
	\begin{equation}
		\label{Eq22}
		\text{Recall}=\frac{T\!P}{T\!P+F\!N},
	\end{equation}
	\begin{equation}
		\label{Eq23}
		\text{WEM}=\frac{\text{IoU}+\text{Precision}+\text{Recall}}{3},
	\end{equation}
	where \(A\), \(B\), \(T\!P\), \(F\!P\), and \(F\!N\) denote the bounding box, ground truth, true positive, false positive, and false negative, respectively. IoU can reveal the matching degree between bounding box and ground truth, which can be seen as the detection effectiveness. Precision refers to the proportion of ground truth among the detected time-frequency position, while recall refers to the proportion of ground truth that is detected among the bounding box. 
\end{enumerate}

\subsection{Simulation Results}

\begin{figure*}[!t]
	\centering
	\subfloat[]{\includegraphics[width=2.3in]{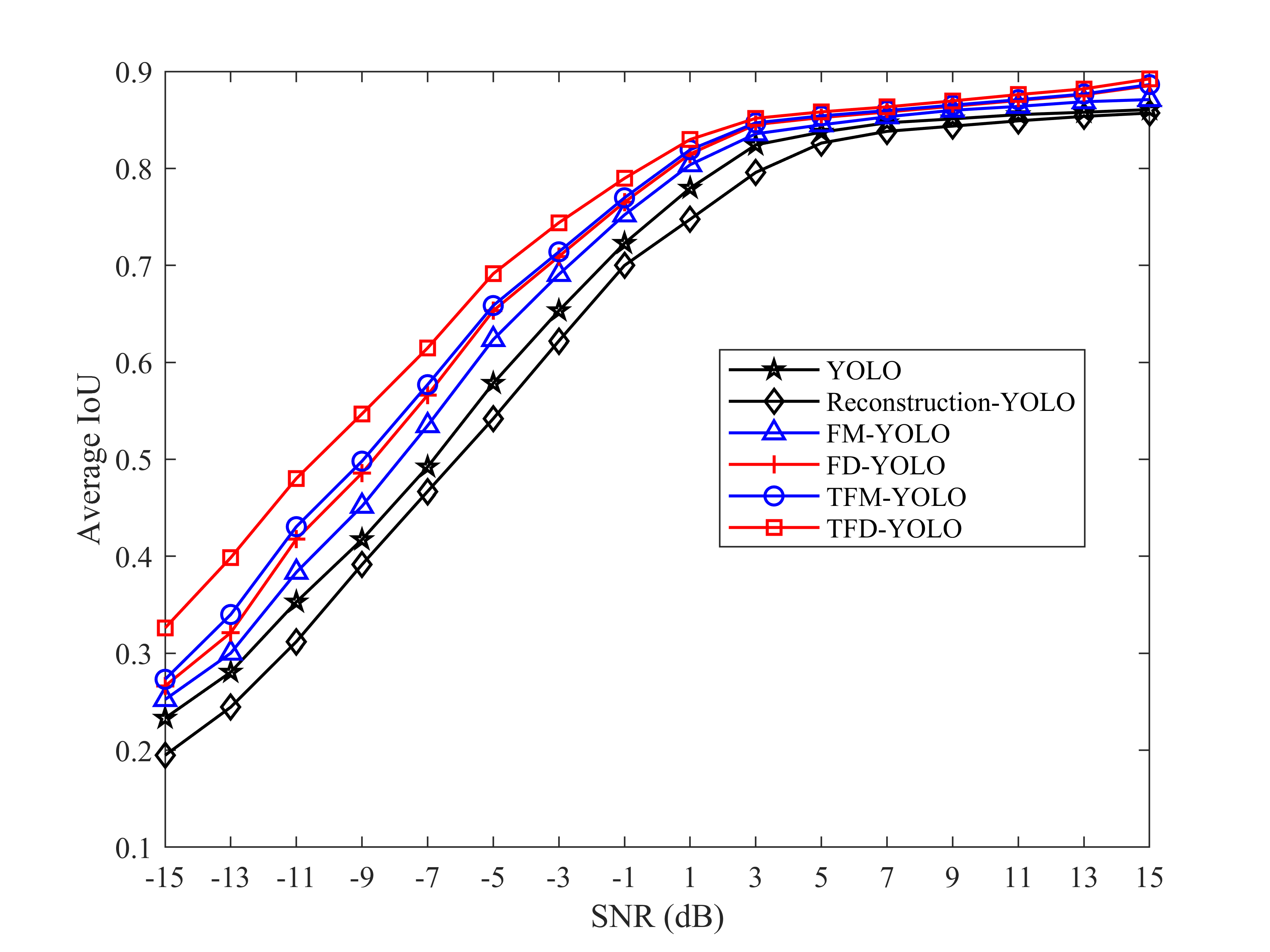}%
		\label{212223_first}}
	\hfil
	\subfloat[]{\includegraphics[width=2.3in]{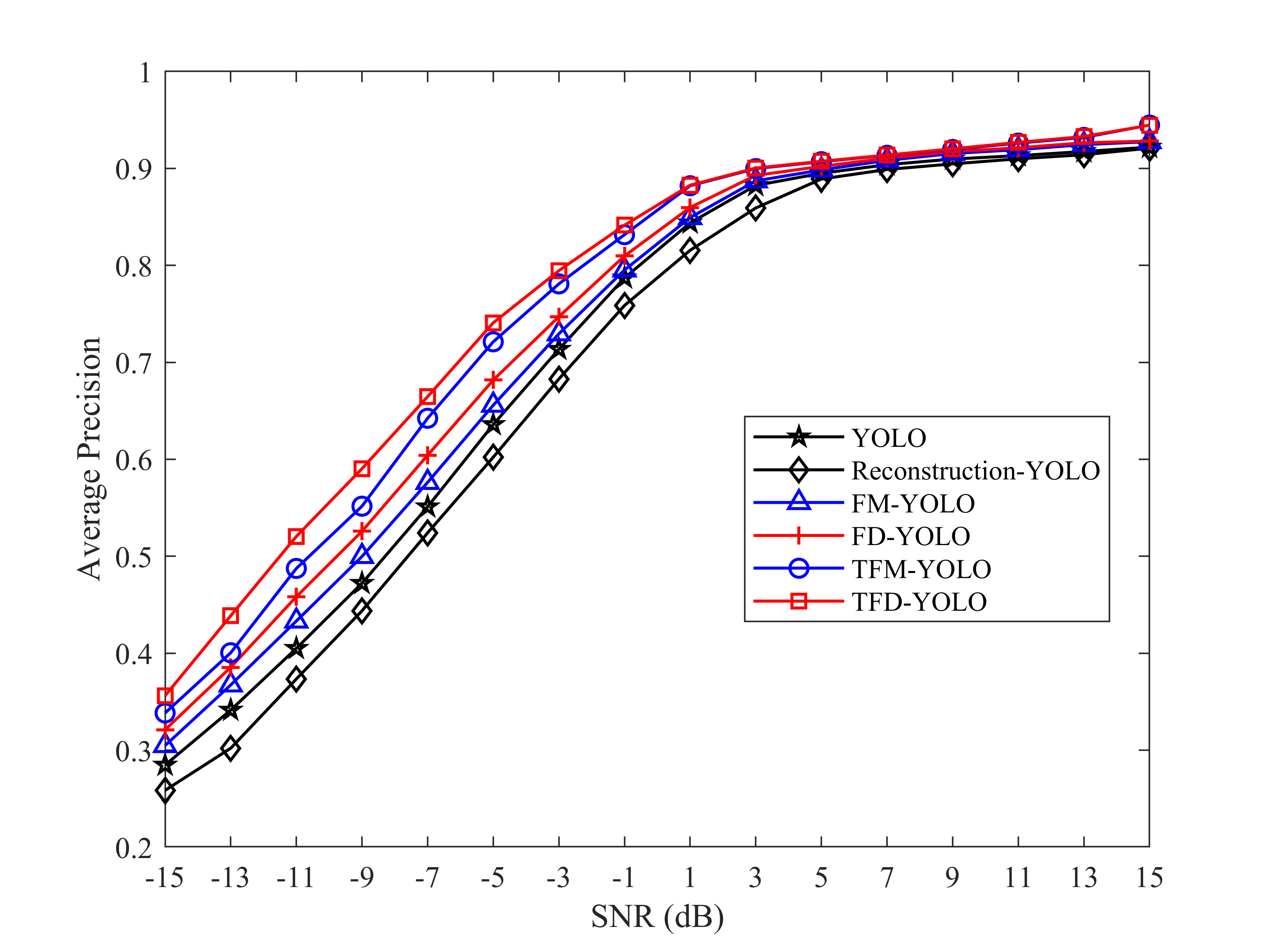}%
		\label{212223_second}}
	\hfil
	\subfloat[]{\includegraphics[width=2.3in]{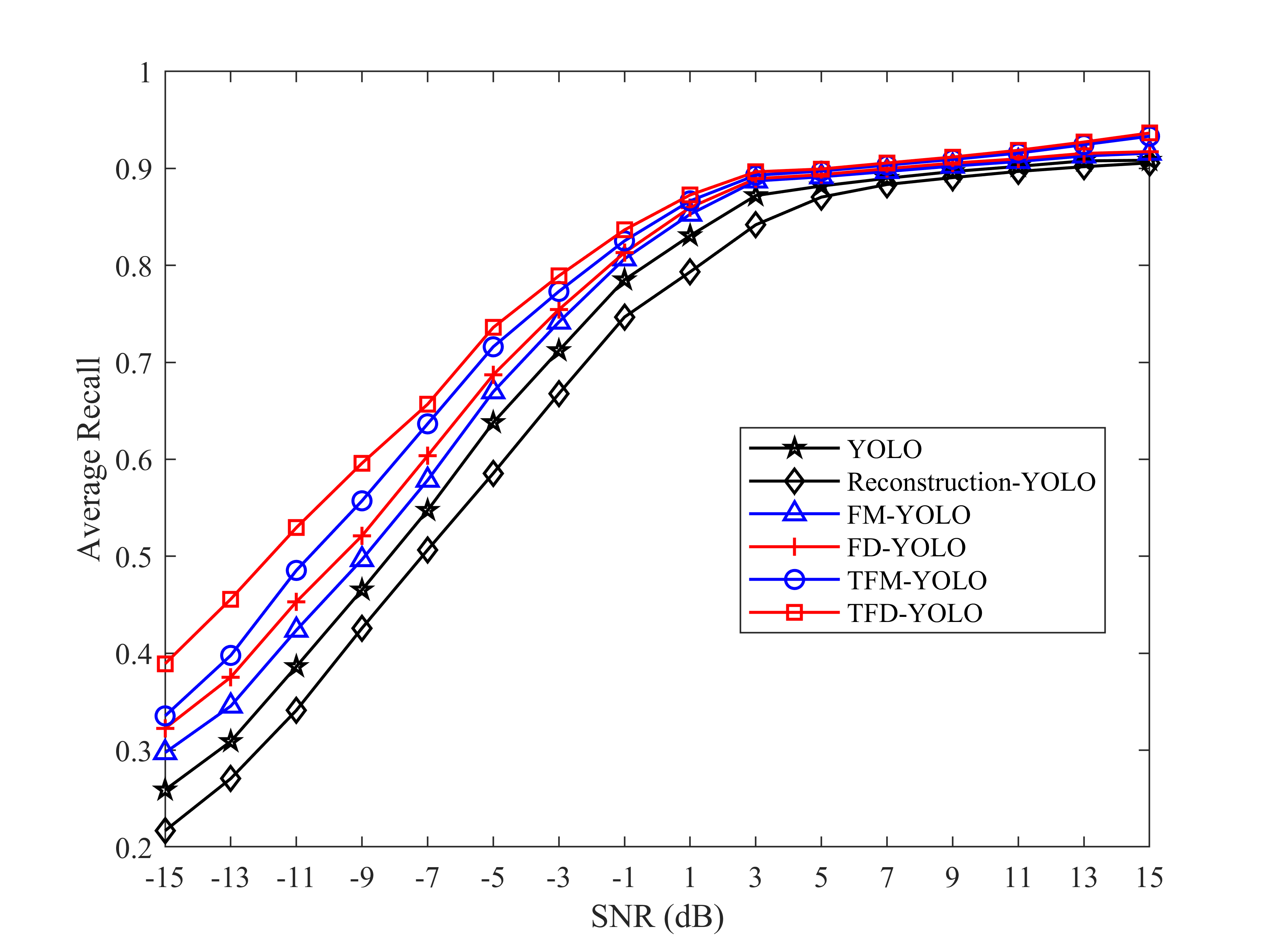}%
		\label{212223_third}}
	\caption{Evaluation metrics of different algorithms versus SNR. (a) Average IoU. (b) Average precision. (c) Average recall.}
	\label{Fig_12}
\end{figure*}

The evaluation metrics of different algorithms are simulated as shown in Fig. \ref{Fig_12}, it can be seen that the performance of all algorithms will improve as SNR increases. However, the metrics of reconstruction-YOLO are lower than those of YOLO. This is because the reconstructed network only performs denoising and interference removal for TFIs at pixel-level, which may not reliably and consistently enhance semantic-level detection accuracy, especially for dim and samll targets locating within a large time-frequency range. Compared to the limited accuracy improvement achieved by the frequency domain parameters correction, the correcting of both time and frequency domain parameters leads to a greater improvement. Especially under low SNR situations, the remarkable correlation properties of ZC sequences can effectively detect the time-frequency position of broadcast frames termed as dim and small targets. Although model-level decoupling can improve detection accuracy to some extent, it can be observed that data-level decoupling achieves better evaluation metrics. This is because the pre-processing of signals based on prior knowledge can be seen as focusing on frequency regions that are likely to contain drone broadcast frames, which significantly improve the learning capability and detection accuracy of the YOLOv7 detector. While both algorithms correct the time-frequency parameters of bounding boxes generated by the YOLOv7 detector, the bounding boxes with data-level decoupling exhibit more precise frequency domain positioning that those with model-level decoupling, resulting in higher average IoU, precision, and recall.

The evaluation metrics for different algorithms are provided in Table \ref{tab:table3}. It can be observed that the proposed algorithm achieves the best detection performance by correcting the time and frequency domain parameters at the data level, which indicates the accuracy and robustness for drone broadcast frames detection termed as dim and small targets. Compared to the model-level decoupling algorithm, the proposed algorithm improves the average IoU, precision, and recall by 3\%, 1.6\%, and 2.4\%, respectively. Besides, despite the bigger processing delay caused by the time-frequency parameters correction operations, the detection latency of 0.0439 s is tolerable for the practical deployment.

\begin{table}[!t]
	\caption{Average Evaluation Metrics of Different Algorithms \label{tab:table3}}
	\centering
	\begin{tabular}{c c c c c}
		\hline
		Algorithm & IoU & Precision & Recall & Detection Latency\\
		\hline
		YOLO & 0.6528 & 0.7111 & 0.6995 & 0.0225 s \\
		Reconstruction-YOLO & 0.6303 & 0.6911 & 0.6716 & 0.0252 s \\
		FM-YOLO & 0.6744 & 0.7245 & 0.7204 & 0.0336 s \\
		FD-YOLO & 0.6907 & 0.7370 & 0.7326 & 0.0343 s \\
		TFM-YOLO & 0.6963 & 0.7549 & 0.7481 & 0.0434 s\\
		TFD-YOLO & \textbf{0.7197} & \textbf{0.7672} & \textbf{0.7661} & \textbf{0.0439} s \\
		\hline
	\end{tabular}
\end{table}

\begin{figure}[!t]
	\centering
	\includegraphics[width=2.5in]{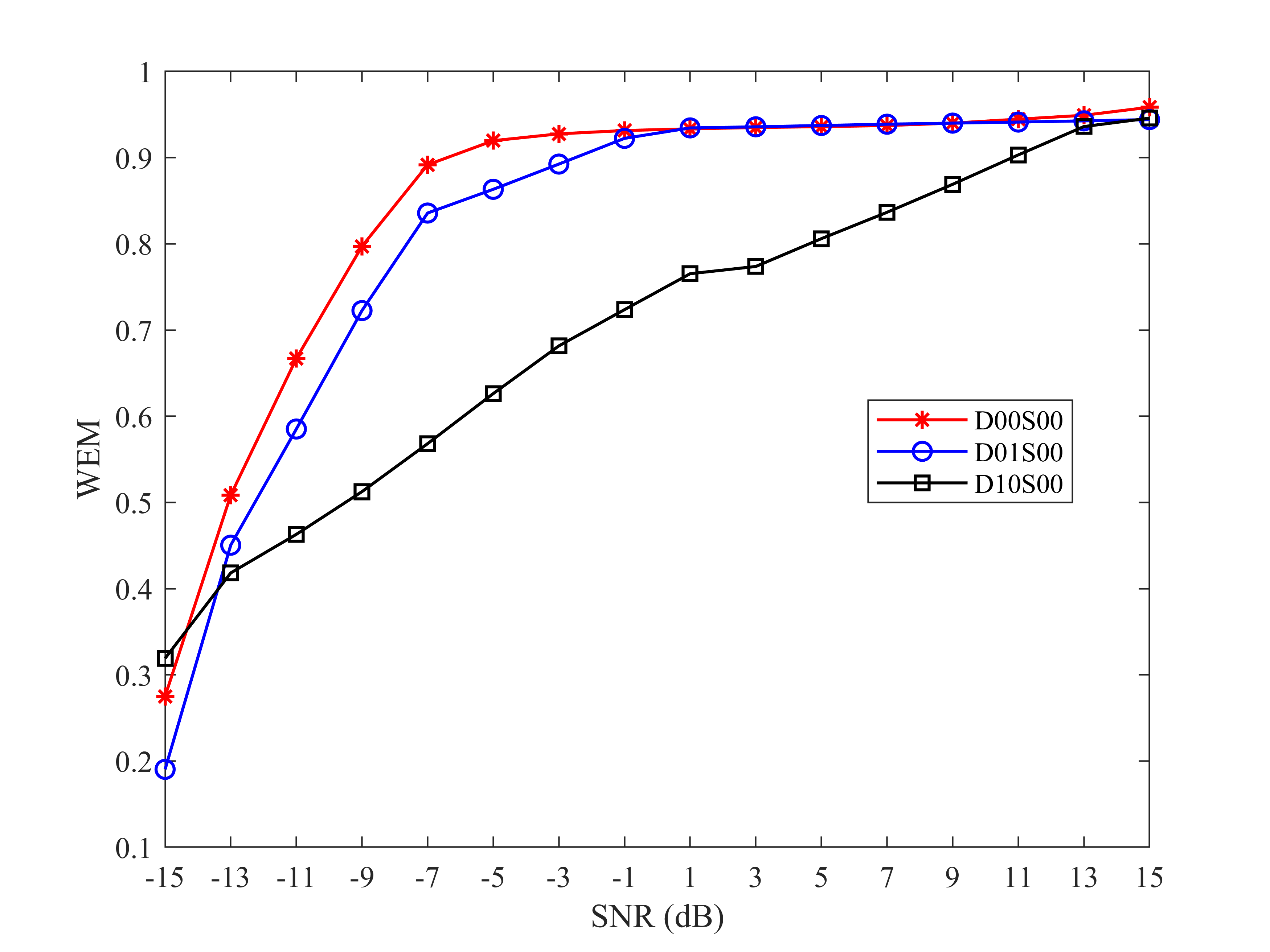}
	\caption{WEM of the proposed algorithm versus drone flight distances.}
	\label{Fig_13}
\end{figure}

Since drones' flight speed can reach 15 m/s, the relative flight distance to the receiver will change rapidly, which make it crucial to evaluate the robustness of the proposed drone broadcast frames detection algorithm under varying flight distance. As illustrated in Fig. \ref{Fig_13}, the proposed algorithm achieves a detection accuracy exceeding 80\% when the flight distance is less than 40 m. Although the average IoU, precision, and recall decrease with increasing flight distance, the proposed algorithm still maintains an detection accuracy of approximately 70\% within the flight distance range of 80\(\sim\)150 m. The simulation results demonstrate the robustness of the proposed algorithm across different flight distances, highlighting its effectiveness in drone broadcast frames detection.

\begin{figure}[!t]
	\centering
	\includegraphics[width=2.5in]{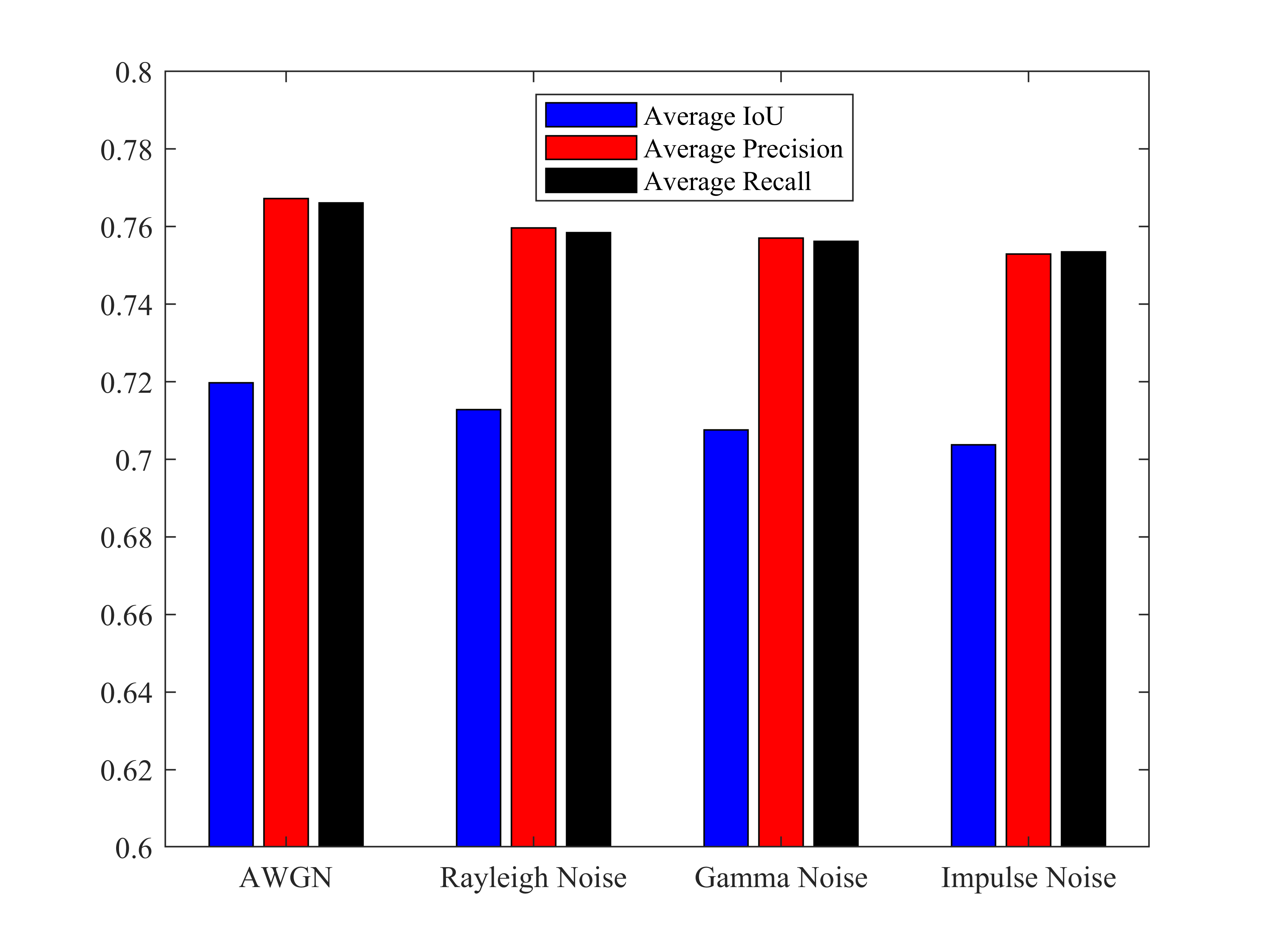}
	\caption{Evaluation metrics of the proposed algorithm versus noise types.}
	\label{Fig_14}
\end{figure}

As mentioned above, to evaluate the robustness of the proposed algorithm under different environmental noise conditions, the weighted evaluation metric is simulated as shown in Fig. \ref{Fig_14}. Since Gaussian noise causes the least signal degradation, the proposed algorithm achieves the highest performance under AWGN. Rayleigh noise has a relatively minor impact on the feature representation of TFIs, while ZC sequences exhibit robustness against Gamma noise, leading to only a slight performance drop. Moreover, the proposed segmented energy refinement algorithm not only eliminates interference signals with high energy intensity but also mitigates the impact of impulse-like high-energy noise, resulting in an average evaluation metrics degradation of only 1.43\%.

\begin{figure}[!t]
	\centering
	\includegraphics[width=2.5in]{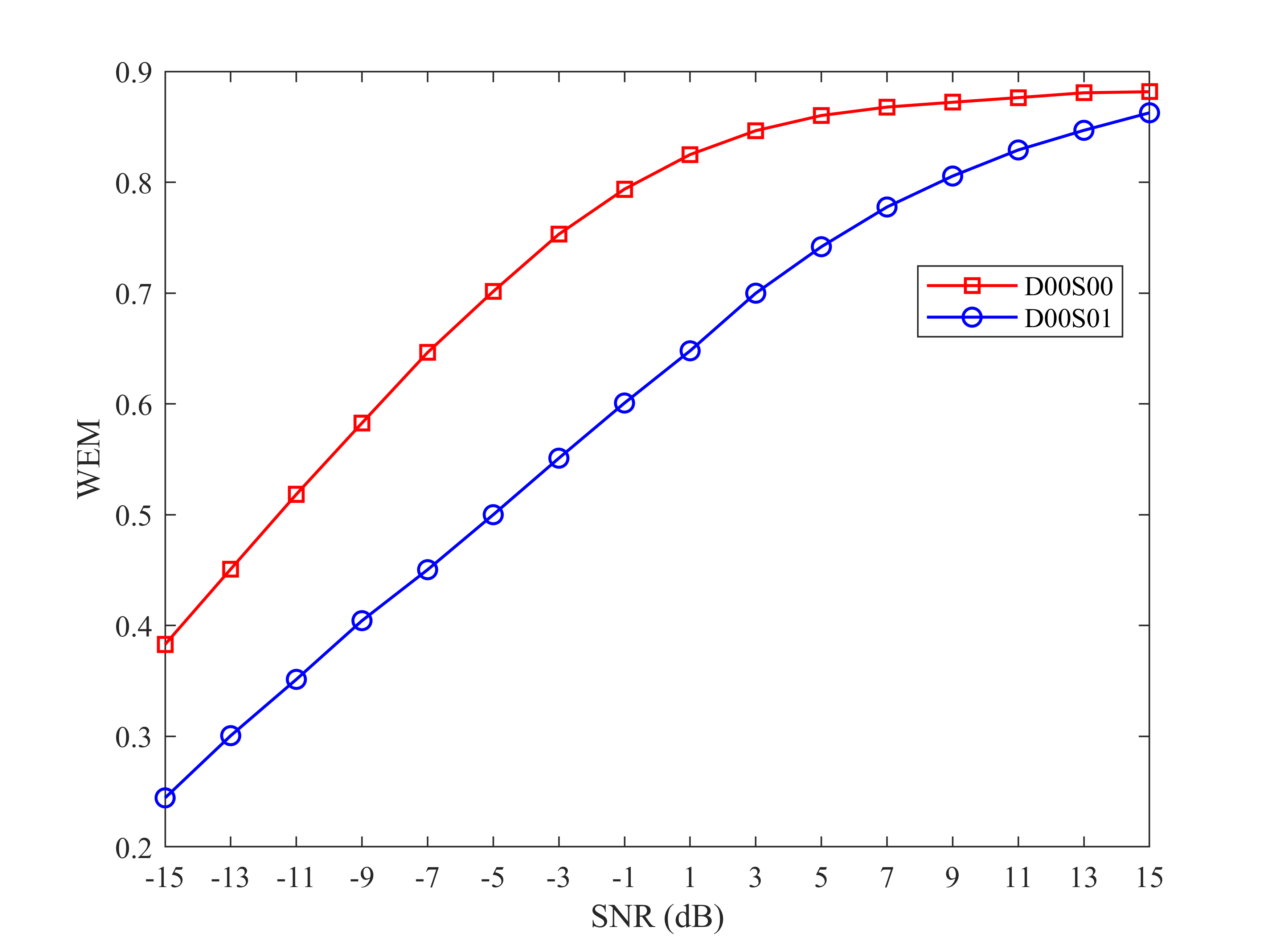}
	\caption{WEM of the proposed algorithm versus flight visual environment.}
	\label{Fig_15}
\end{figure}

The weighted evaluation metric under different flight visual environments is simulated as shown in Fig. \ref{Fig_15}. It can be seen that the received drone signal strength is lower in non-LOS environment, leading to an average performance degradation of 13.29\% compared to the LoS scenario. However, as the channel conditions improve, the detection accuracy of the proposed algorithm increases, which can achieve a detection accuracy of over 80\% under favorable channel condition even in non-LOS environment.

\begin{figure*}[!t]
	\centering
	\subfloat[]{\includegraphics[width=2.3in]{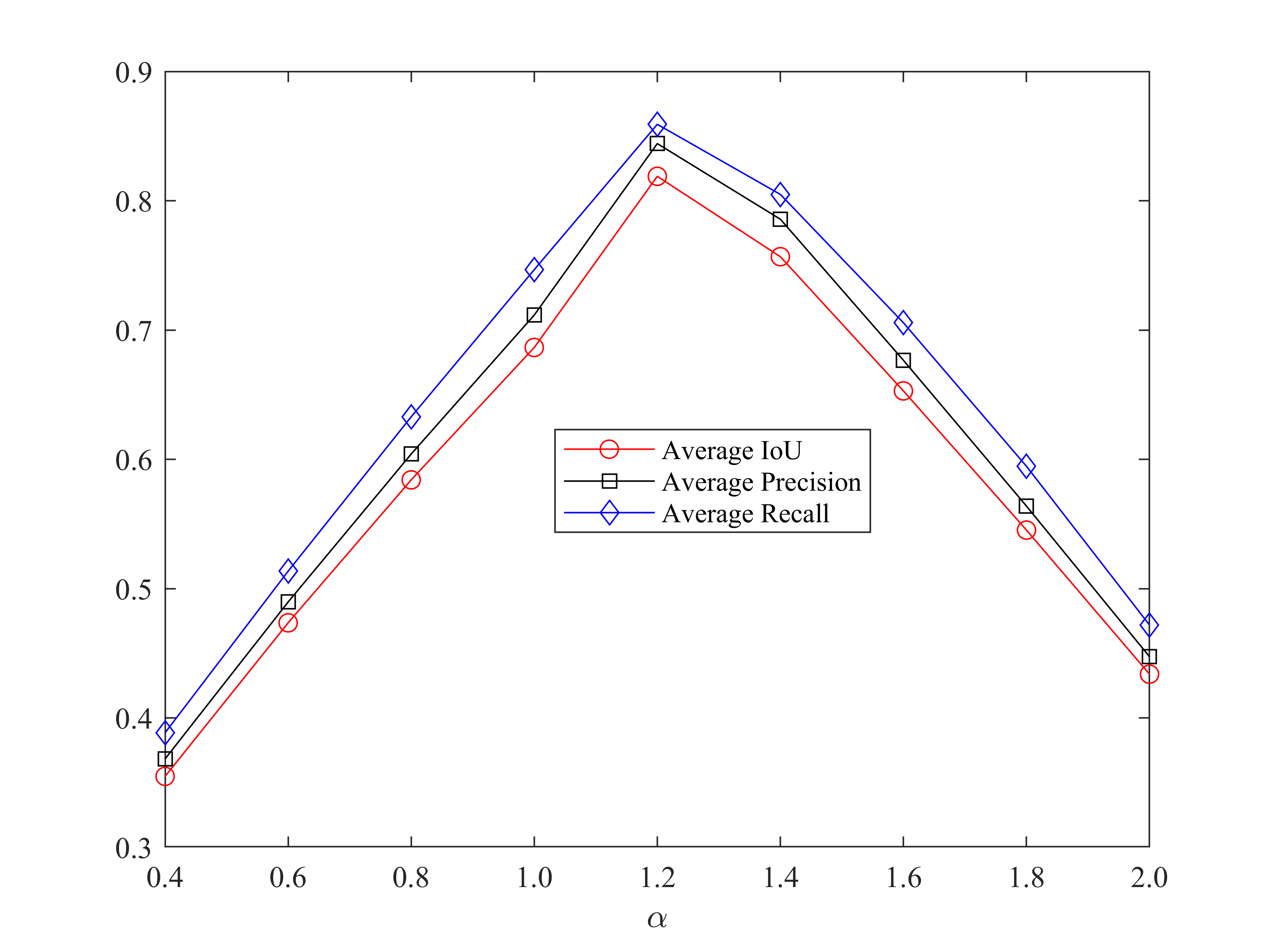}%
		\label{272829_first}}
	\hfil
	\subfloat[]{\includegraphics[width=2.3in]{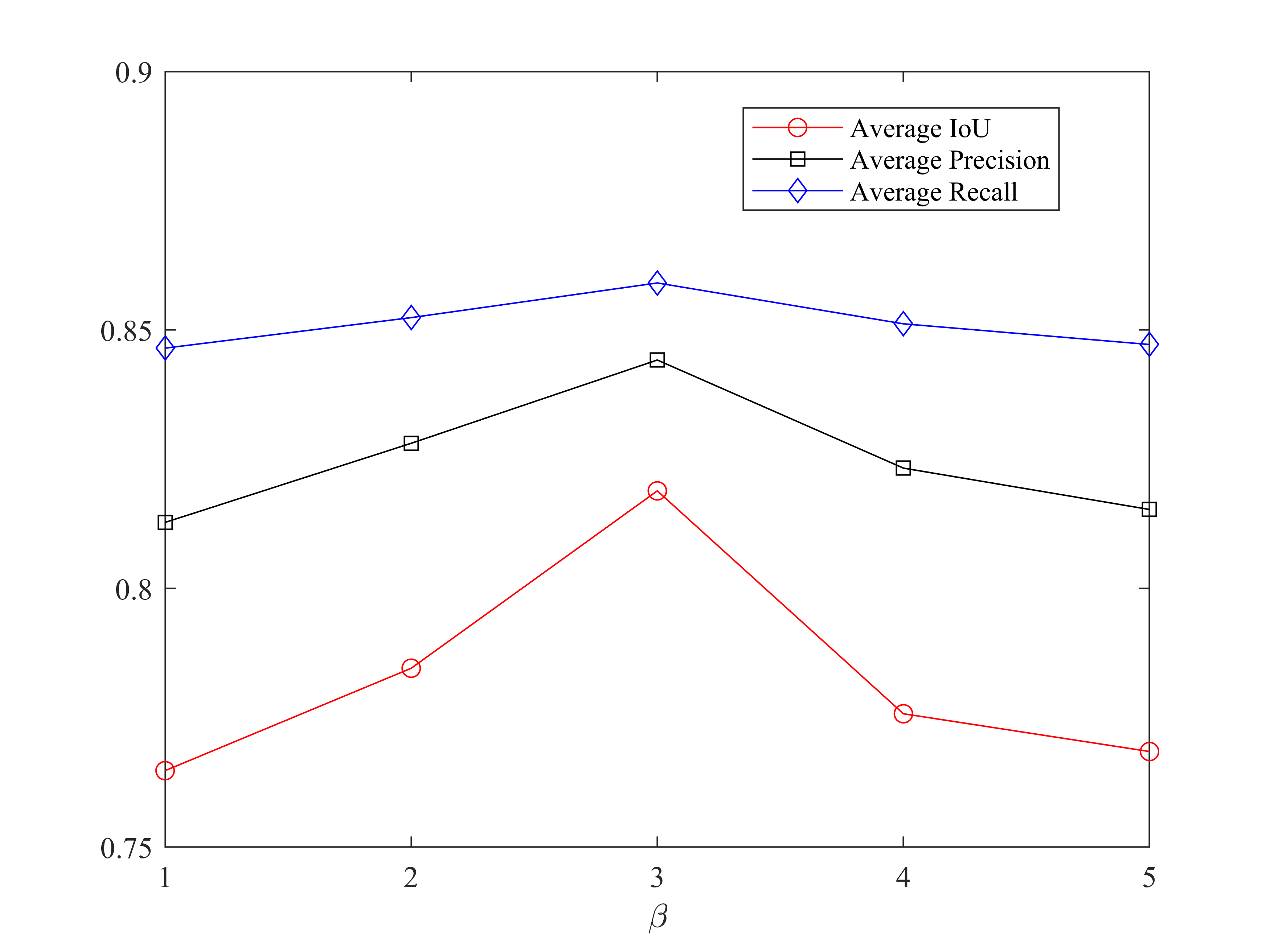}%
		\label{272829_second}}
	\hfil
	\subfloat[]{\includegraphics[width=2.3in]{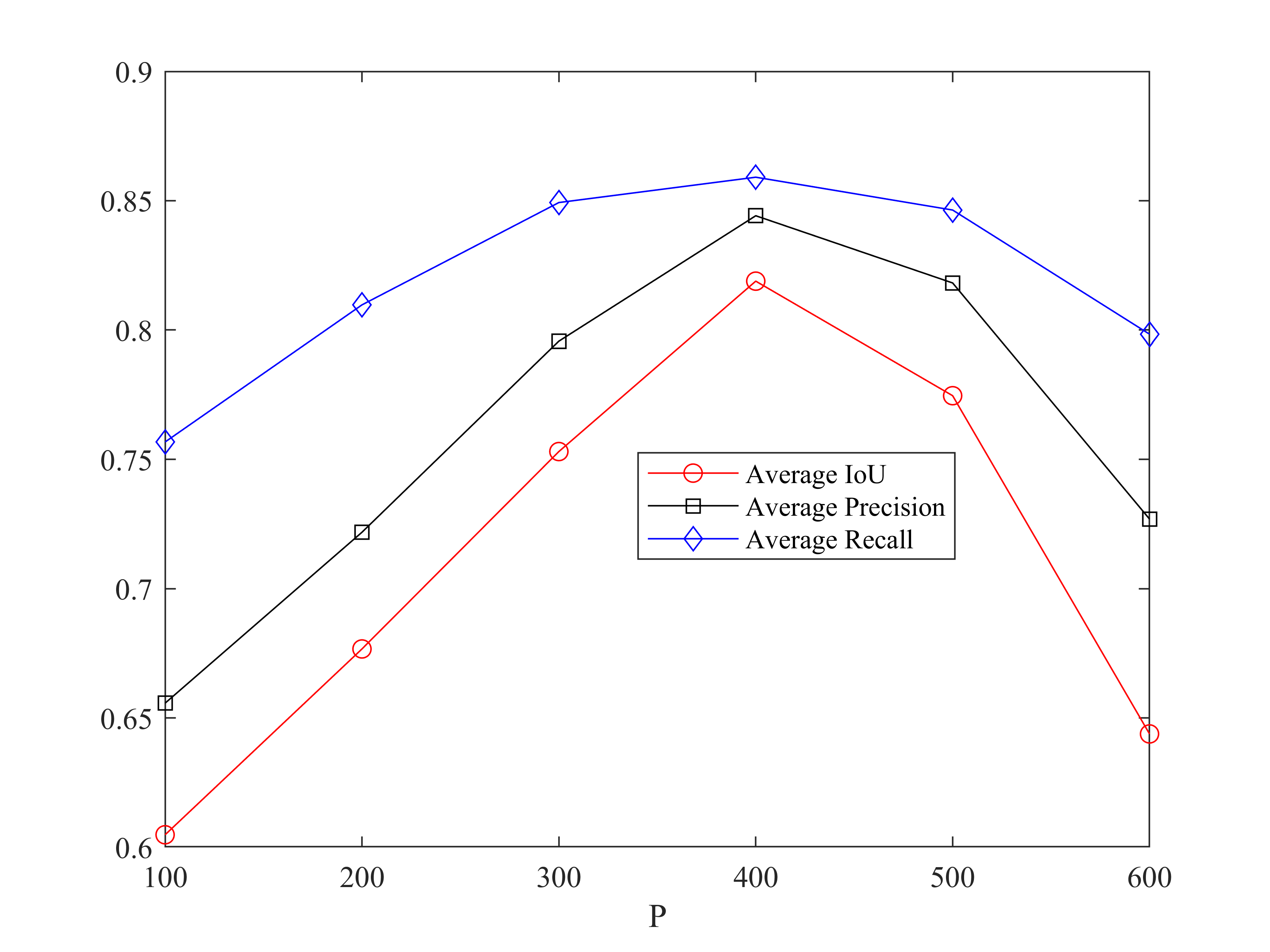}%
		\label{272829_third}}
	\caption{Evaluation metrics of segmented energy refinement method versus parameter selection. (a) \(\tiny{\alpha}\). (b) \(\tiny{\beta}\). (c) \(\tiny{P}\).}
	\label{Fig_16}
\end{figure*}

The impact of three key parameters in the proposed segmented energy refinement method on drone broadcast frames detection accuracy is illustrated in Fig. \ref{Fig_16}. The relative energy intensity of drone broadcast frame, interference signals, and noise varies with different SNR situations, which subsequently affects the intensity of \(\gamma(m)\) and \(E_Q\). Therefore, the selection of weighting factor \(\alpha\) plays a crucial role on the construction of index set \(I\). It can be seen that as \(\alpha\) varies from 0.4 to 2.0, the average IoU, precision, and recall first increase and then decrease, with reaching the peaks at \(\alpha=1.2\). This is because the segment of \(\gamma(m)\) containing the ZC sequences may be refined when \(\alpha<1.2\), which causes \(m^*\) to be located at non-broadcast frames' time domain positions. Conversely, certain segment of \(\gamma(m)\) corresponding to interference signals with high energy strength may be retained when \(\alpha>1.2\), which causes \(m^*\) to be located at the time domain positions of interference signals. \(\beta\) is utilized to add consecutive \(q\) to avoid the omitted refinement of interference signals with high energy strength, and the best detection accuracy is achieved when \(\beta=3\).
Fig. \ref{272829_third} indicates that the optimal value of \(P\) of the proposed segmented energy refinement method should be 400. The shorter \(P\) tends to result in higher \(E(q)\) for the segment containing ZC sequences when \(P<400\), which will lead to the refinement of \(\gamma(m)\) corresponding to ZC sequences. Conversely, the longer \(P\) will cause the \(\gamma(m)\) of certain interference signals to be averaged out by noise, resulting in the retention of \(\gamma(m)\) corresponding to interference signals and wrong correction of time domain parameters. 

\begin{figure}[!t]
	\centering
	\includegraphics[width=2.5in]{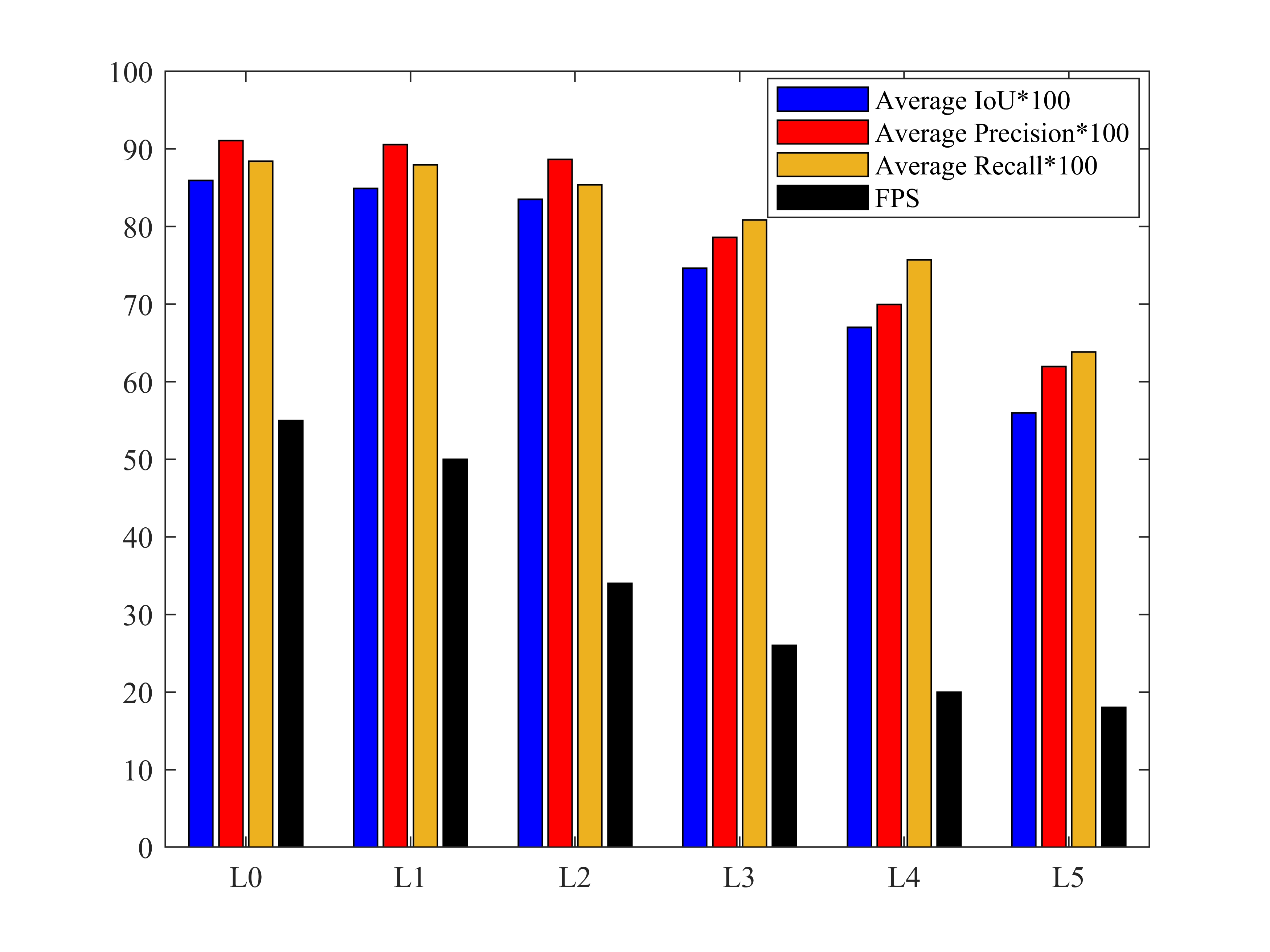}
	\caption{Detection accuracy and speed versus sampling durations.}
	\label{Fig_17}
\end{figure}

In addition to high detection accuracy requirement, drone regulation demands a high response speed of algorithms to meet the real-time regulation requirement. Existing works primarily evaluate the real-time performance of drone detection algorithms from two aspects like FPS and latency. Fig. \ref{Fig_17} illustrates that by transforming RF signals with different sampling duration into TFIs, the proposed drone broadcast frames detection algorithm exhibits variations in both detection accuracy and FPS. Specifically, the drone broadcast frames occupy a larger region in TFIs when sampling duration L0 = 1 ms, and the proposed algorithm only needs to consider the weak targets detection under low SNR situations. Although the cost of time domain parameters correction is minimal and FPS achieves the highest in this case, the proposed algorithm can only process RF signals with a duration of 55 ms, which meets the detection requirement of 20 FPS\textcolor{blue}{\cite{ref45}} but leads to a high detection latency. The detection accuracy and FPS both decrease as sampling duration increases, while FPS reaches the minimum value of 18 when L5 = 50 ms. Although the detection requirement of 20 FPS cannot be met due to the high cost of time parameters correction for broadcast frames termed as dim and small targets in TFIs, the per-second detection latency is only 100 ms, which is acceptable. Since the proposed drone broadcast frames detection algorithm is designed as a auxiliary and front module for the object tracking and trajectory prediction of drones, the detection and classification latency in \textcolor{blue}{\cite{ref11}},\textcolor{blue}{\cite{ref12}}, and\textcolor{blue}{\cite{ref26}} can be compared with that of\textcolor{blue}{\cite{ref16}}. The latency of the proposed drone RID algorithm based on ZC sequences and TFIs is 100 ms, which is much less than 4.5 s in\textcolor{blue}{\cite{ref26}} even with the addition of algorithm processing time. 

\begin{figure}[!t]
	\centering
	\includegraphics[width=2.5in]{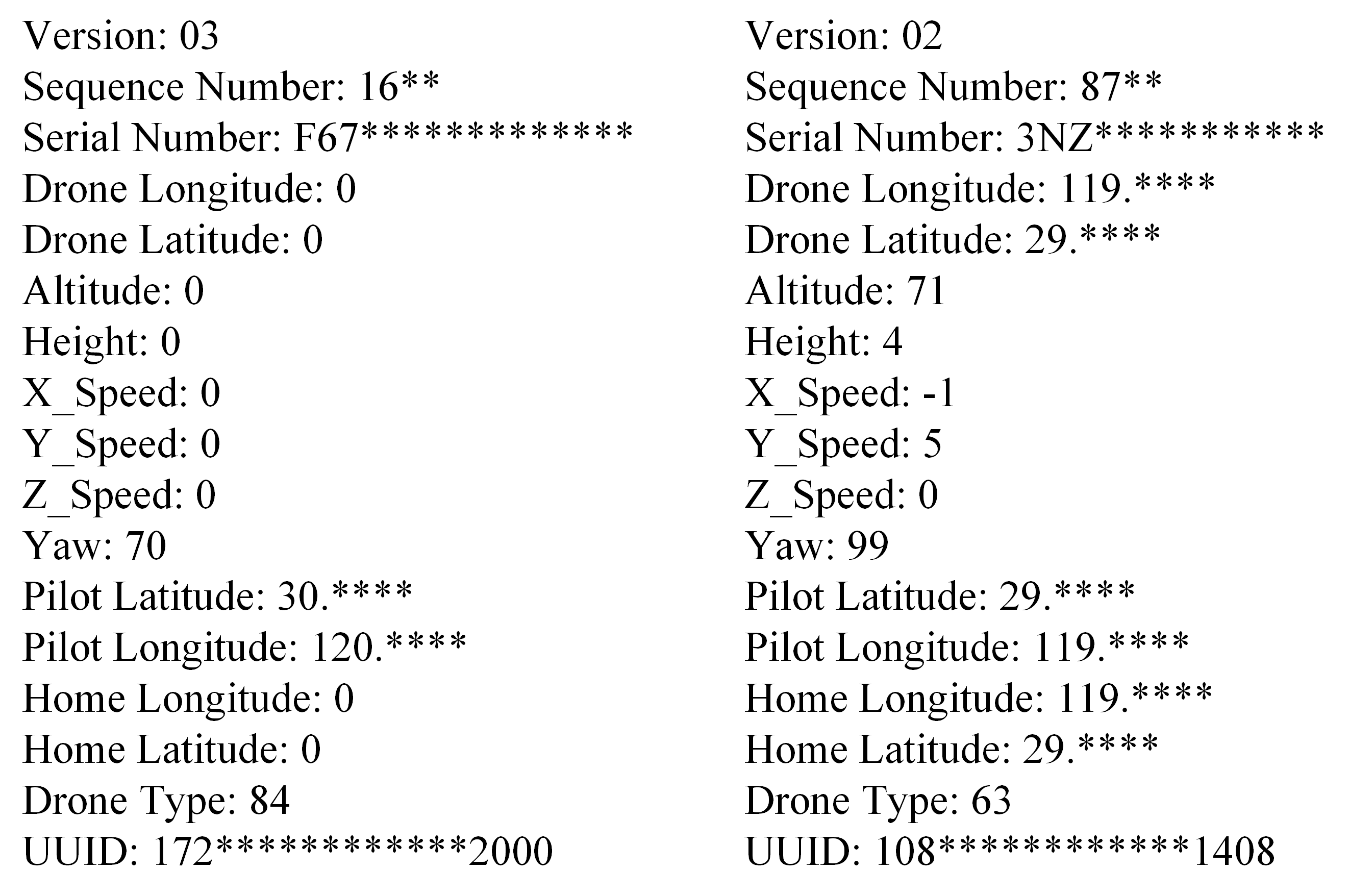}
	\caption{Payload information in two broadcast frames from different datasets.}
	\label{Fig_18}
\end{figure}

The partial payload information of two broadcast frames from different datasets is shown in Fig. \ref{Fig_18}, and some privacy information is masked. It can be observed that by decoding the broadcast frames, information such as the speed, position, and serial number of drone and pilot can be obtained. Among them, the serial number and the universally unique identifier (UUID) are both uniquely associated with an individual, which provides a key technical solution for drone RID. In addition, 37 broadcast frames from a total of 14 drones of 10 categories in Table \ref{tab:table1} are decoded. Only 1 frame failed in CRC, i.e., the decoding failed, which means the identification, association, and positioning of drone and pilot is achieved with an accuracy of 97.30\% based on the proposed algorithm in this paper.

\section{Conclusion}
\label{sec6}
This paper proposed a dim and small target detection algorithm for broadcast frames to achieve effective drone RID. The prior knowledge of modulation parameters and frame structures was established. Transmission frequency and signal bandwidth were leveraged to correct the frequency domain parameters of bounding boxes, while ZC sequences and frame length were utilized to correct the time domain detection parameters of broadcast frames. Besides, a segmented energy refinement method was applied to mitigate time domain parameter deviation caused by interference signals with high energy strength. Since the sampling duration determined the occupied regions of broadcast frame and whether it corresponded to a small target, the trade-off between detection accuracy and speed of the proposed algorithm was established to meet different detection requirements. For the detected broadcast frames, a decoding algorithm was proposed to recover the essential payload information for RID. Simulation results demonstrated that the proposed algorithms outperformed existing algorithms in terms of both detection accuracy and speed. And the proposed algorithm also maintained robust performance across varying flight distances, different flight visual environment, and diverse types of environment noise. The decoding results also indicated that the effective of utilizing broadcast frames to achieve RID.

\end{document}